\documentclass[aps,tightenlines,twocolumn,prd,nofootinbib,
superscriptaddress,showpacs,preprintnumbers,eqsecnum,floatfix]{revtex4-2}
\def\bea{\begin{eqnarray}}
\def\be{\begin{equation}}
\def\ee{\end{equation}}
\def\eea{\end{eqnarray}}
\def\bal{\begin{align}}
\def\eal{\end{align}}

\usepackage[usenames,dvipsnames]{color}
\usepackage{graphics}
\usepackage{graphicx}
\usepackage{amsmath}
\usepackage{amssymb}
\usepackage{bbold}
\usepackage{bm}
\usepackage{slashed}
\usepackage{float}
\usepackage{psfrag}
\usepackage{footnote} 
\usepackage{xcolor}
\usepackage{textcomp}
\usepackage{cancel}
\usepackage{dcolumn}
\begin{document}

\title{Simple high-accuracy method for solving bound-state equations with the Cornell potential in momentum space }

\author{Alfred Stadler}
\email{stadler@uevora.pt}
\affiliation{Departamento de F\'isica, Universidade de \'Evora, 7000-671 \'Evora, Portugal}
\affiliation{Laborat\'orio de Instrumenta\c{c}\~ao e F\'isica Experimental de Part\'iculas---LIP, Avenida Professor Gama Pinto, 2, 1649-003 Lisboa, Portugal}
\affiliation{Departamento de F\'isica, Instituto Superior T\'ecnico, Universidade de Lisboa, Avenida Rovisco Pais, 1, 1049-001 Lisboa, Portugal}

\author{Elmar P. Biernat}
\email{elmar.biernat@tecnico.ulisboa.pt}
\affiliation{C2TN and Departamento de Engenharia e Ci\^encias Nucleares, Instituto Superior T\'ecnico, Universidade de Lisboa, Campus Tecnol\'ogico e Nuclear, 2695-066 Bobadela, Portugal}
\affiliation{Laborat\'orio de Instrumenta\c{c}\~ao e F\'isica Experimental de Part\'iculas---LIP, Avenida Professor Gama Pinto, 2, 1649-003 Lisboa, Portugal}
\affiliation{
CFTP, Instituto Superior T\'ecnico, Universidade de Lisboa, Avenida Rovisco Pais, 1, 
1049-001 Lisboa, Portugal}
\affiliation{Departamento de F\'isica, Instituto Superior T\'ecnico, Universidade de Lisboa, Avenida Rovisco Pais, 1, 1049-001 Lisboa, Portugal}

\author{Vasco Valverde}
\email{v.valverde@campus.fct.unl.pt}
\affiliation{Universidade Nova de Lisboa, Faculdade de Ci\^encias e Tecnologia, Campus de Caparica, 2829-516 Caparica, Portugal}

\date{\today}
 \begin{abstract}
The well-known Cornell quark-antiquark potential in momentum space contains singularities both in its one-gluon-exchange (OGE) and linear confining parts, which prevents a direct use of the convenient Nystr\"om method to solve the corresponding bound-state integral equation for the meson masses. While it has been known for a long time how the Coulomb-type singularity in the OGE potential can be treated with a subtraction technique, only very complicated methods have been developed to deal with the stronger singularity in the linear potential. In this work, we present a simple subtraction method to remove this singularity from the kernel, such that the Nystr\"om method becomes applicable. Derivatives of the wave function, that appear as a result of the subtraction, are represented by means of interpolating functions, for which we found Lagrange polynomials to be very efficient. Test calculations show excellent agreement with exactly known energy eigenvalues. By increasing the number of integration points and the order of the Lagrange interpolation polynomials, extremely high accuracy can be achieved. This method can also be extended to relativistic Bethe-Salpeter type equations with singular kernels.
 \end{abstract}


\maketitle



\section{Introduction}\label{sec1}
The nonrelativistic Cornell potential model~\cite{PhysRevLett.34.369,Eichten:1978,Eichten:1980}, and \lq\lq relativized'' variations of it~\cite{Godfrey:1985aa,PhysRevD.72.054026}, have been very successful in describing the masses and decay rates of heavy quarkonia. They represent the interaction between heavy quarks and antiquarks as a combination of a color-Coulomb potential created by one-gluon exchange which dominates at small distances, and a linear confining potential of nonperturbative origin that is mainly responsible for the behavior of the interaction at longer distances. Usually, a nonrelativistic Schr\"odinger equation with this potential is solved in coordinate space~\cite{Soni:2017wvy,Soni:2020tji,Kang:2006jd}. Lowest-order relativistic corrections are often added, of which the spin-dependent contributions yield spin-spin, spin-orbit, and tensor interactions~\cite{PhysRevD.23.2724}. Spin-independent corrections, which contain nonlocal terms, are usually ignored.

Generally formulated within Schr\"odinger theory, such models do not account for the dynamical covariant structure of quarks. However, for mesons with at least one light quark a relativistic treatment is necessary, and in addition, the implementation of dynamical mass generation and spontaneous chiral symmetry breaking is indispensable for a realistic description of light mesons, in particular the pions. This can be achieved in nonperturbative approaches based on quantum field theory, such as the Dyson-Schwinger/Bethe-Salpeter  formalism~\cite{Eichmann:2016yit, Alkofer:2000wg, Fischer:2006ub, Roberts:1994dr, Roberts:2000aa, Maris:2003vk} and the Covariant Spectator Theory (CST)~\cite{Gross:1969rv, Biernat:2014a, Biernat:2014xaa, Savkli:1999me,Gross:1991te,Gross:1991pk, Gross:1994he, Gross:1964mla, Gross:1964zz}. The CST formalism also allows for the inclusion of a covariant generalization of the Cornell potential, such that the CST bound-state equations for mesons in the nonrelativistic limit reduce again to the Schr\"odinger equation with the Cornell potential. 

These manifestly covariant relativistic approaches are represented in momentum space, which seems to be the only practical way to take relativity exactly into account, and they are formulated and solved in the form of integral equations.
However, in momentum space, the Cornell potential contains singularities, which greatly complicate the application of standard numerical solution techniques. In particular, the very convenient and often used Nystr\"om method \cite{Nystrom:1930aa,Atkinson:1997fv} is incompatible with the presence of singularities in the kernel of the integral equation. 

These singularities are of the same type in the nonrelativistic Cornell potential and in its relativistic generalization in CST. It is therefore sufficient to look for solution methods of the simpler nonrelativistic problem first and then apply them to the relativistic case.

Several works have addressed the problem of the nonrelativistic Cornell potential in momentum-space. How the Coulomb part of the potential can be treated through a subtraction technique to be compatible with the Nystr\"om method has already been shown in \cite{Kwon:1978}. Unfortunately, no similar subtraction method was found to work with the Nystr\"om method for the more difficult linear potential. 

Alternatively, in Ref.~\cite{Andreev:2017kie}, the linear-confining problem in S-waves was solved using a quadrature method where the singularity of the integral is incorporated into the weight functions. This method gives remarkable high-precision results, but is very involved in practice, because the weight functions must be adapted specifically for each integral appearing in the Schr\"odinger equation.  

References~\cite{Chen:2012sv,Chen:2013hna} used a screened and therefore nonsingular Cornell potential, which made it possible to solve the momentum-space Schr\"odinger equation with the Nystr\"om method and an extended Simpson's rule and other quadrature rules. However, when the unscreened limit is approached, the numerical accuracy worsens significantly.

Others~\cite{Eyre:1986,Norbury:1992jv,Maung:1993aa,Hersbach:1993xz,Tang:2001ii,Deloff:2007} opted for applying Galerkin methods \cite{Atkinson:1997fv}, in which the unknown wave function is represented as a linear combination of basis functions, and the solution of the resulting linear eigenvalue problem yields the corresponding expansion coefficients, together with the bound-state energies. The singularities in the kernel are integrable---a principal value singularity in the case of the linear potential---and can be integrated numerically with the Sloan method, i.e., by placing quadrature points symmetrically around the singularity.

For this method to work well in practice, one needs to carefully choose appropriate basis functions that reproduce the correct asymptotic behavior of the wave function for small and large momenta. This behavior is determined by the potential and also depends on the partial wave. If the kernel is modified in any way, for instance by including form factors or terms of relativistic origin, the asymptotic behavior may change, and the basis functions have to be modified accordingly. What accuracy can be achieved ultimately depends on an appropriate choice of the basis.

It is in fact also possible to remove the principal-value singularity of the linear potential altogether by a subtraction technique~\cite{Deloff:2007,Leitao:2014jha}, however, at the expense of introducing the first derivative of the wave function under the integral. This seemed to exclude the Nystr\"om method, but represents no obstacle for the application of a Galerkin method. In Ref.~\cite{Leitao:2014jha}, we have successfully used this method with a basis of B-spline functions. 

Nevertheless, in view of the already mentioned challenges connected with an adequate choice of basis functions in the Galerkin method, in this work we revisit the subtracted form of the linear potential problem, and demonstrate how it can be made amenable for the Nystr\"om method, thereby presenting a simpler, more flexible and efficient numerical method for solving the Cornell potential in momentum space. 
The new procedure can provide extremely accurate results, and can also be applied to the fully relativistic linear confining problem in the CST equation, because the singularity structure remains unchanged under its covariant generalization. 

In Sec.~\ref{sec2} we first review the partial wave form of the momentum-space Schr\"odinger equation with the Cornell potential and its singularity structure, as well as the basic ideas of the Nystr\"om method. Then we show how the Coulomb singularity can be made compatible with the Nystr\"om method, and why the same technique does not seem to work for the linear potential. We go on to demonstrate our solution to this problem, which consists of representing the first and second derivatives of the wave function, that appear as a consequence of the subtraction of the singularity, in terms of interpolating functions. Section~\ref{sec:coll} displays the resulting discretized version of the momentum-space Schr\"odinger equation in detail and brings it into matrix-equation form. Our numerical results are presented and discussed in Sec.~\ref{sec:num}, and finally Sec.~\ref{sec:concl} gives a summary of this work and our conclusions.

\section{The Cornell potential in momentum space}\label{sec2}
The Cornell quark-antiquark potential~\cite{PhysRevLett.34.369,Eichten:1978,Eichten:1980} in coordinate space is composed of a Coulomb-like short-range interaction, originating from one-gluon-exchange, and a linearly rising confining potential, of the form
\begin{equation}
\tilde{V}({\bf r})= -\frac{\alpha}{r}+\sigma r \, ,
\end{equation}
where $r=|{\bf r}|$ is the distance between the two quarks, $\alpha$ is the strong coupling constant, including a color factor, and the constant $\sigma$ characterizes the strength of the confining potential.

The Fourier-transform of the Coulomb potential
\begin{equation}
\tilde{V}_C({\bf r})= -\frac{\alpha}{r} \, 
\end{equation}
 is very well known,
\begin{equation}
\label{eq:VCq}
V_{C}({\bf q}) 
 =\int d^{3}r \tilde{V}_C({\bf r}) e^{i{\bf q}\cdot {\bf r}}
 =-\frac{4\pi\alpha}{{\bf q}^2} \, , 
\end{equation}
which is obtained by first transforming a screened version of it and then taking the unscreened limit. The momentum transfer ${\bf q}={\bf k}-{\bf p}$ is the difference between the relative momenta in the initial and final state.

In order to find the momentum-space version of the linear potential, 
\begin{equation}
\tilde V_L({\bf r})=\sigma r \, ,
\end{equation}
one can apply the same general idea, but different screening methods can be used. In~\cite{Leitao:2014jha}, we found that starting from a particular kind of screening leads to the convenient representation
\begin{equation}
V_L({\bf q}) = \left[ V_A({\bf q}) -(2\pi)^3\delta^{(3)}({\bf q})\int \frac{d^3 q'}{(2\pi)^3} V_A({\bf q}') \right] \, ,
\label{eq:VL}
\end{equation}
where 
\begin{equation}
V_A({\bf q}) = - \frac{8\pi\sigma}{{\bf q}^4} \, .
\label{eq:VA}
\end{equation}
The form of the ``potential'' (\ref{eq:VL}) is somewhat peculiar and should be interpreted as a distribution. It will always be multiplied with a wave function and integrated over the loop momentum $\bf k$, which produces meaningful expressions.  

This becomes clear when the momentum-space Cornell potential,
\begin{equation}
V({\bf q})=V_C({\bf q})+V_L({\bf q}) \, ,
\end{equation}
 is now inserted into the Schr\"odinger equation for the momentum-space wave function $\psi ({\bf p} )$ of a two-body system with reduced mass $m_R$,
\begin{multline}
\frac{p^2}{2m_R}  \psi ({\bf p} )
+  
\int  \frac{d^3 k}{(2\pi)^3} V_C ({\bf p}-{\bf k})
\psi ({\bf k} )
\\ 
+  
\mathrm {P}\!\!\!  \int  \!\!\!\!   \frac{d^3 k}{(2\pi)^3} V_A ({\bf p}-{\bf k})
 \left[   \psi ({\bf k} )
 -   \psi ({\bf p} )\right] =
 E   \psi ({\bf p} )\, .
 \label{eq:SE}
\end{multline} 
Both integrands are singular  at ${\bf k}={\bf p}$, and it appears that the singularity in  $V_A$ is even much stronger than the one in $V_C$. However, it is weakened by the wave-function subtraction in the numerator that arises from the Dirac-delta term in $V_L$ of Eq.~(\ref{eq:VL}). As a result, the integral over $V_A$ in (\ref{eq:SE}) reduces to a Cauchy principal value integral, indicated by the symbol $\mathrm {P}\!\! \int $, and is therefore well defined~\cite{Leitao:2014jha}.

Next, we expand Eq.~(\ref{eq:SE}) into partial waves. Introducing the unit vectors $\hat{\bf k}$ and $\hat{\bf p}$, such that ${\bf p}=p \hat{\bf p}$  and ${\bf k}=k \hat{\bf k}$, and abbreviating $x=\hat{\bf k}\cdot \hat{\bf p}$, the potentials, depending only on ${\bf q}^2=p^2+k^2-2pkx$, can be written as functions of $p$, $k$, and $x$. Let us first look at the Coulomb potential. The angular dependence is 
represented as a series of Legendre polynomials $P_\ell(x)$,
\begin{equation}
V_C(p,k,x) =  \sum_{\ell =0}^\infty \frac{4\pi}{2\ell+1} V_{C,\ell}(p,k)P_\ell(x) \, .
\end{equation}
The expansion coefficients are determined through the angular integral
\begin{align}
\label{eq:VC-PW}
V_{C,\ell}(p,k)
 &=2\pi 
 \int_{-1}^1 dx P_\ell(x) V_C(p,k,x) \nonumber\\
 &=-8\pi^2 \alpha \int_{-1}^1 dx \frac{P_\ell(x)} {p^2+k^2-2pkx} \nonumber\\
 & = -\frac{8\pi^2 \alpha}{pk} Q_\ell(y) 
 \, ,
\end{align}
where the Legendre functions of the second kind, $Q_\ell(y)$, depend on
\begin{equation}
y=\frac{p^2+k^2}{2pk} \, ,
\end{equation}
and are singular at $y=1$, which occurs when $k=p$. 
The representation
\begin{equation}
Q_\ell(y) = P_\ell(y)Q_0(y)-W_{\ell-1}(y) \, ,
\end{equation}
with
\begin{equation}
W_{\ell-1}(y) = \sum_{m=1}^\ell \frac{1}{m} P_{\ell-m}(y) P_{m-1}(y) \, ,
\end{equation}
is very useful here.
It shows that the singularities in the potential matrix elements for all angular momenta have their origin in $Q_0(y)$, which can also be written
\begin{equation}
Q_0(y) = \frac{1}{2} \ln \left| \frac{y+1}{y-1} \right| 
= \frac{1}{2} \ln \left( \frac{p+k}{p-k}\right)^2 \, ,
\end{equation}
whereas $W_{\ell-1}(y)$ is not singular and contributes only for $\ell \ge 1$. 

The partial-wave decomposition of the linear potential $V_L({\bf q})$ of (\ref{eq:VL}) has been derived in~\cite{Leitao:2014jha}. Because of its nature as a distribution, it is easiest to write how it acts on the partial-wave components $\psi_\ell(p)$ of the wave function\footnote{Because of rotational symmetry, neither the partial-wave potentials nor wave functions depend on the eigenvalue of the $z$-component of the angular momentum, which is therefore suppressed throughout.}. We find
\begin{multline}
\label{eq:VL-PW}
 \int_0^\infty  \frac{d k k^2}{(2\pi)^3}
V_{L,\ell}(p,k)\psi_\ell(k) \\
= 
\int_0^\infty  \frac{d k k^2}{(2\pi)^3}
\left[
V_{A,\ell}(p,k)  \psi_\ell(k) 
  - 
V_{A,0}(p,k)  \psi_\ell(p)
 \right] \, ,
\end{multline}
where the partial-wave elements of $V_A$ are
\begin{align}
&V_{A,\ell}(p,k) \nonumber\\
&=2\pi 
 \int_{-1}^1 dx P_\ell(x) V_A(p,k,x) \nonumber\\
& =
\frac{8\pi^2\sigma}{(pk)^2}
\left[ 
 P_\ell(y) Q'_0(y)
 +
 P'_\ell(y) Q_0(y) 
 -W'_{\ell-1}(y) \right] 
\nonumber \\
& = -\frac{8\pi^2\sigma}{(pk)^2}
\Biggl[ 
\left(\frac{2pk }{ p^2-k^2} \right)^2 P_\ell(y)
-
P'_\ell(y) Q_0(y)
+
W'_{\ell-1}(y)
\Biggr] .
\label{eq:VAPWfinal}
\end{align}
In addition to the singularity in $Q_0(y)$, that is also present in the Coulomb potential (\ref{eq:VC-PW}), the first term
is even more strongly singular and the source of all the difficulties in this calculation.

We can now write the partial-wave Schr\"odinger equation for the Cornell potential as
\begin{widetext}
\begin{multline}
\frac{p^2}{2m_R} 
\psi_\ell(p)
 -\frac{\alpha}{\pi p }
 \int_0^\infty  dk \, k
 \left[ P_\ell(y)Q_0(y)-W_{\ell-1}(y)\right]
\psi_\ell(k) 
\\
 -
 \frac{\sigma}{\pi p^2}
 \mathrm {P}\!\!\! \int_0^\infty dk 
 \left\{
 \frac{4p^2 k^2}{(k^2-p^2)^2}
 \left[ P_\ell(y) \psi_\ell(k)-\psi_\ell(p) \right] 
 -
Q_0(y)
 P'_\ell(y)\psi_\ell(k)
 +W'_{\ell-1}(y)\psi_\ell(k)
 \right\}
 = E \psi_\ell(p) \, .
 \label{eq:SEl}
\end{multline}
 
\end{widetext}
Next we turn our attention to the problem of solving this equation numerically.

\subsection{The Nystr\"om method}
\label{sec:Nym}
The Schr\"odinger equation in momentum space (\ref{eq:SEl}) is a homogeneous Fredholm integral equation of the second kind, of the form
\begin{equation}
\label{eq:Fred}
\int_a^b dk\, K(p,k) \phi(k)=E \phi(p) \, ,
\end{equation}
where $K(p,k)$ is its kernel. One convenient method to solve it numerically is the Nystr\"om method~\cite{Nystrom:1930aa}, which is based on the numerical integration of a function $f(k)$ by means of a quadrature rule,
\begin{equation}
\label{eq:Quad}
\int_a^b  f(k) dk \approx \sum_{j=1}^N w_j f(k_j) \, .
\end{equation}
The quadrature points $k_j$ and weights $w_j$ are determined by the choice of the quadrature method and the integration interval, which in our case extends from $0$ to $\infty$. After applying (\ref{eq:Quad}) to (\ref{eq:Fred}), one obtains a closed set of $N$ linear equations if for the external variable $p$ we choose each of the points $\{k_j\}$ of the quadrature rule,
\begin{equation}
\sum_{j=1}^N w_j K(k_i,k_j) \phi(k_j)=E \phi(k_i) \, .
\end{equation}
The continuous variables $p$ and $k$ have thereby been discretized to take on only values from the same set, and they are now essentially only distinguished by their indices. From here on we will refer to the set of quadrature points as $\{p_i\}$ and use the notation $p_i$ or $p_j$ rather than $k_i$ or $k_j$.

If we introduce the notation $\phi_i = \phi(p_i)$ and $M_{ij}=w_j K(p_i,p_j)$, the discretized Schr\"odinger equation can be written as a linear eigenvalue problem for the $N\times N$ matrix $M_{ij}$, 
\begin{equation}
\label{eq:NyM}
\sum_{j=1}^N M_{ij} \phi_j=E \phi_i \, ,
\end{equation}
where the eigenvector for each eigenvalue $E$ is a set of values of the corresponding eigenfunction at the quadrature points, $\phi(p_i)$. 

\subsection{Subtraction of logarithmic singularities from the kernel}
Several terms in the kernel of Eq.\ (\ref{eq:SEl}) are singular when $k=p$ (and $y=1$). Therefore, the diagonal elements $M_{ii}$ in (\ref{eq:NyM}) do not exist and the Nystr\"om method cannot be applied. 

However, by means of a trick due to Land\'e~\cite{Lande+Kwon}, the logarithmic singularities in the $Q_0(y)$ terms can be eliminated.
Taking advantage of the well-known result~\cite{Kwon:1978}
\begin{equation}
\int_0^\infty dk \frac{Q_0(y)}{k}=\frac{\pi^2}{2} \, ,
\label{eq:intQ0y}
\end{equation}
and using $P_\ell(1)=1$, we can write
\begin{align}
\int_0^\infty dk\, & k P_\ell(y) Q_0(y) \psi_\ell(k)  \nonumber\\
& = \int_0^\infty dk\, \left[ k P_\ell(y) Q_0(y)\psi_\ell(k)-\frac{p^2}{k}Q_0(y)\psi_\ell(p) \right]
\nonumber\\
&\hspace{20pt}+  \int_0^\infty dk\,  \frac{p^2}{k}Q_0(y) \psi_\ell(p)\nonumber\\
&=
p \int_0^\infty dk\, \left[ \frac kp P_\ell(y) Q_0(y)\psi_\ell(k)-\frac{p}{k}Q_0(y)\psi_\ell(p) \right] 
\nonumber\\
&\hspace{20pt}+ p^2 \frac{\pi^2}{2}\psi_\ell(p) \, .
\end{align}
The integrand in square brackets vanishes in the limit $k\to p$ and therefore does not contribute to the diagonal of the matrix $M$. 
The complete Coulomb term in the Schr\"odinger equation (\ref{eq:SEl}) can now be written 
\begin{multline}
\label{eq:VC-subtr}
 -\frac{\alpha}{\pi p }
 \int_0^\infty  dk \, k
 \left[ P_\ell(y)Q_0(y)-W_{\ell-1}(y)\right]
\psi_\ell(k) \\
=  -\frac{\alpha}{\pi p }
 \int_0^\infty  dk \,  
 \left[ k\,  Q_\ell(y)\psi_\ell(k)-\frac{p^2}{k}Q_0(y)\psi_\ell(p) \right] \\
-\frac{\alpha \pi p}{2} \psi_\ell(p) \, .
\end{multline}
For the Nystr\"om method, we need to know the limit of the integrand of the second line of (\ref{eq:VC-subtr}) as $k \to p$, which comes only from the $W_{\ell-1}(y)$ term hidden in $Q_\ell(y)$,
\begin{equation}
\label{eq:VClk=p}
\lim_{k\to p} \left[k\,   Q_\ell(y)\psi_\ell(k)-\frac{p^2}{k}Q_0(y)\psi_\ell(p)\right] = -p W_{\ell-1}(1) \psi_\ell(p) \, ,
\end{equation}
and it is easy to see that 
\begin{equation}
W_{\ell-1}(1) =  \sum_{m=1}^\ell \frac{1}{m} \, .
\end{equation}

Using the Land\'e subtraction technique, the pure Coulomb problem can therefore be solved quite easily with the Nystr\"om method.
 
A similar singular term as in the Coulomb potential appears also in the kernel (\ref{eq:VAPWfinal}) of the linear potential in all partial waves with $\ell \ge 1$. We can treat it with a slight variation of the same idea,
\begin{align}
\label{eq:intQ0lin}
 & \int_0^\infty  dk\, Q_0(y) P'_\ell(y)\psi_\ell(k) \nonumber\\
 & = \int_0^\infty  dk\, \left[ Q_0(y) P'_\ell(y)\psi_\ell(k) - \frac pk Q_0(y) P'_\ell(1)\psi_\ell(p) \right] \nonumber\\
 &\hspace{20pt} +\int_0^\infty  dk\,  \frac pk Q_0(y) P'_\ell(1)\psi_\ell(p)  \nonumber\\
 & = \int_0^\infty  dk\, \left[ Q_0(y) P'_\ell(y)\psi_\ell(k) - \frac pk Q_0(y) P'_\ell(1)\psi_\ell(p) \right] \nonumber\\
 &\hspace{20pt} +p \frac{\pi^2}{2} P'_\ell(1)\psi_\ell(p) \, ,
\end{align}
where 
$P'_\ell(1)=\ell (\ell+1)/2$.

Again, the integrand vanishes when $k\to p$ and can therefore be omitted from the diagonal elements of the matrix $M$ when the Nystr\"om method is used. There is no need to write $\mathrm {P}\!\! \int_0^\infty $ in the first line of (\ref{eq:intQ0lin}) because this is now a regular integral.

\subsection{Subtraction of the principal value singularities}

There is only one singular term remaining in the kernel of the linear potential in (\ref{eq:SEl}), but this is the most problematic one. After pulling out a factor $2p^2$ for convenience, we have to deal with
\begin{equation}
\label{eq:linsing}
 \mathrm {P}\!\!\! \int_0^\infty dk 
 \frac{2 k^2}{(k^2-p^2)^2}
 \left[ P_\ell(y) \psi_\ell(k)-\psi_\ell(p) \right] \, .
\end{equation}

 There appears to be a double pole at $k=p$, but the numerator goes to zero like $(k-p)$,  so it is actually only a single pole, and the principal value integral exists.
 
To see this, we expand the factor in brackets in the numerator of (\ref{eq:linsing}) in a Taylor series around $k=p$,
\begin{multline}
 P_\ell(y) \psi_\ell(k)-\psi_\ell(p) = (k-p) \psi'_\ell(p) \\
 +\frac{ (k-p)^2}{2!}  
 \left[  \frac{P'_\ell(1)}{p^2} \psi_\ell(p) +\psi''_\ell(p)
\right]
+\frac{(k-p)^3}{3!}  R_\ell(k) \, ,
 \label{eq:Taylor1}
\end{multline}
where the function $R_\ell(k)$ is the remainder of the Taylor series  after the terms up to second order have been subtracted and $(k-p)^3/3!$ has been factored out. The only important property of $R_\ell(k)$ in this context is that it is finite at $k=p$. 
Equation (\ref{eq:Taylor1}) confirms that one power of $k-p$ in the denominator of (\ref{eq:linsing}) is canceled and that the singularity is therefore only a single pole.

Substituting (\ref{eq:Taylor1}) into (\ref{eq:linsing}) shows that the integrand in the vicinity of $k=p$ behaves like
\begin{multline}
 \label{eq:intsub}
 \frac{2k^2}{(k^2-p^2)^2}
 \left[ P_\ell(y) \psi_\ell(k)-\psi_\ell(p) \right] \\
 =
 \frac{2k^2}{k+p} \frac{ \psi'_\ell(p)}{k^2-p^2} 
 +  \frac{k^2}{(k+p)^2}  \left[  \frac{P'_\ell(1)}{p^2} \psi_\ell(p) +\psi''_\ell(p)
\right] \\
+\frac{k^2(k-p)}{3(k+p)^2}R_\ell(k) \\
 =  \frac{p\psi'_\ell(p)}{k^2-p^2} +\frac{2k+p}{(k+p)^2} \psi'_\ell(p)\\
 +  \frac{k^2}{(k+p)^2}  \left[  \frac{P'_\ell(1)}{p^2} \psi_\ell(p) +\psi''_\ell(p)
\right]
+\frac{k^2(k-p)}{3(k+p)^2}R_\ell(k) \, .
\end{multline}

For the Nystr\"om method to work, this singularity has to be removed. 
In \cite{Leitao:2014jha} we have shown that this can be done with another Land\'e subtraction. The reason why we have separated the term proportional to $\psi'_\ell(p)$ in (\ref{eq:intsub}) into two parts is that, when the first one is subtracted, \begin{widetext}
\begin{equation}
\label{eq:linsing2}
 \mathrm {P}\!\!\! \int_0^\infty dk 
 \frac{2 k^2}{(k^2-p^2)^2}
 \left[ P_\ell(y) \psi_\ell(k)-\psi_\ell(p) \right] 
=
 \int_0^\infty dk \left\{
 \frac{2 k^2}{(k^2-p^2)^2}
 \left[ P_\ell(y) \psi_\ell(k)-\psi_\ell(p) \right] -  \frac{p\psi'_\ell(p)}{k^2-p^2} \right\}
 +p \psi'_\ell(p)\mathrm{P}\!\! \int_0^\infty \frac{dk}{k^2-p^2} 
 \, ,
\end{equation}
the well-known integral
\begin{equation}
\mathrm{P}\!\! \int_0^\infty \frac{dk}{k^2-p^2}=0 
\end{equation}
takes care of the principal value singularity. The final result is an ordinary integral, free of singularities,
\begin{equation}
\label{eq:linsing3}
 \mathrm {P}\!\!\! \int_0^\infty dk 
 \frac{2 k^2}{(k^2-p^2)^2}
 \left[ P_\ell(y) \psi_\ell(k)-\psi_\ell(p) \right] 
 = 
 \int_0^\infty dk \left\{
 \frac{2 k^2}{(k^2-p^2)^2}
 \left[ P_\ell(y) \psi_\ell(k)-\psi_\ell(p) \right] -  \frac{p\psi'_\ell(p)}{k^2-p^2} \right\}
 \, .
\end{equation}
The price to pay for this simplification is that the derivative of the wave function makes its entrance into the integrand, which does not seem compatible with the Nystr\"om method. In \cite{Leitao:2014jha} we therefore followed a different strategy to solve the---now singularity-free---integral equation for the Cornell potential, 
\begin{multline}
\left(\frac{p^2}{2m_R} 
-\frac{\alpha \pi p}{2} 
+\frac{\sigma \pi}{2p}P'_\ell(1)\right) \psi_\ell(p)
 -\frac{\alpha}{\pi p }
 \int_0^\infty  dk \,  
 \left[ k\,  Q_\ell(y)\psi_\ell(k)-\frac{p^2}{k}Q_0(y)\psi_\ell(p) \right] \\
 -
 \frac{2\sigma}{\pi}
 \int_0^\infty dk 
 \left\{
 \frac{2 k^2}{(k^2-p^2)^2}
 \left[ P_\ell(y) \psi_\ell(k)-\psi_\ell(p) \right] 
 -  \frac{p\psi'_\ell(p)}{k^2-p^2}
 -
\frac{Q_0(y)}{2p^2}
\left[ 
 P'_\ell(y)\psi_\ell(k) - \frac{p}{k} 
 P'_\ell(1)\psi_\ell(p)
 \right]
 +\frac{W'_{\ell-1}(y)}{2p^2}\psi_\ell(k)
 \right\} \\
 = E \psi_\ell(p) \, ,
 \label{eq:SElnosing}
\end{multline}
 
\end{widetext}
namely to expand $\psi_\ell(p)$ into a set of basis functions whose derivatives can be easily calculated. In \cite{Leitao:2014jha} we chose a basis of modified cubic B-spline functions. 

In this Galerkin method, one solves for the expansion coefficients instead of the function values at grid points. The method works well and gives accurate results, but the choice of basis functions requires special care. It is especially important that the basis is able to accurately reproduce the behavior of the solutions at very small and very large momenta. This asymptotic behavior needs to be determined before the equation is solved, which is not always easy to do. It is one of the advantages of the Nystr\"om method that no advance knowledge of the properties of the solutions is needed.

\subsection{The Nystr\"om method for the subtracted kernels}
The purpose of this work is to show how the Nystr\"om method can be applied to solve the integral equation (\ref{eq:SElnosing}). As already mentioned before, one of the requirements is to be able to calculate the kernel along the diagonal $k=p$.  
That the kernel is not singular at $k=p$ is a necessary, but not a sufficient condition. In the case of the subtracted Coulomb potential, (\ref{eq:intQ0lin}) provides these matrix elements. We need to do the same for the subtracted linear potential.

From (\ref{eq:intsub}) we can read off
\begin{multline}
\label{eq:VAlk=p}
\lim_{k\to p} \left\{
 \frac{2 k^2}{(k^2-p^2)^2}
 \left[ P_\ell(y) \psi_\ell(k)-\psi_\ell(p) \right] -  \frac{p\psi'_\ell(p)}{k^2-p^2} \right\} \\
 = \frac{3}{4p} \psi'_\ell(p) + \frac{1}{4}  \left[  \frac{P'_\ell(1)}{p^2} \psi_\ell(p) +\psi''_\ell(p)
\right] \, .
\end{multline}
The $Q_0(y)$ term of the linear potential in  (\ref{eq:SElnosing}) vanishes in this limit, so only the unproblematic $W'_{\ell -1}(1)\psi_\ell(p)/2p^2 $ has to be added to (\ref{eq:VAlk=p}).

At this point, the situation seems to be even worse than expected from looking at (\ref{eq:SElnosing}), because we need to know the first \emph{and second} derivatives of the unknown function $\psi_\ell(p)$ to construct the complete kernel matrix along the diagonal.

The new idea of this work is to calculate the derivatives of the wave function at the grid points $p_i$ by means of interpolating functions that depend linearly on the values of the wave function at the grid points  in the vicinity of $p_i$. The Schr\"odinger equation can thereby be cast into the form (\ref{eq:NyM}) and solved by standard numerical methods to determine the eigenvalues and eigenvectors of square matrices.

\section{The discretized form of the momentum-space Schr\"odinger equation}
\label{sec:coll}
\noindent 
As already outlined in Sec.\ \ref{sec:Nym}, we discretize the integrations over $k$ according to (\ref{eq:Quad}), rename the grid points $k_j \to p_j$, and choose the points $p$ on the same grid of points $\{p_i\}$. Furthermore, we introduce the abbreviations $\psi_i=\psi_\ell(p_i)$ and $y_{ij}=(p_i^2+p_j^2)/(2p_ip_j)$. With these conventions, the Schr\"odinger equation (\ref{eq:SElnosing}) becomes
\begin{widetext}
\begin{multline}
\left(\frac{p_i^2}{2m_R} 
-\frac{\alpha \pi p_i}{2} 
+\frac{\sigma \pi}{2p_i}P'_\ell(1)\right) \psi_i
 -\frac{\alpha}{\pi p_i }
\sum_{j \ne i} w_j \,  
 \left[ p_j\,  Q_\ell(y_{ij})\psi_j-\frac{p_i^2}{p_j}Q_0(y_{ij})\psi_i \right] 
 + \frac{\alpha}{\pi} w_i W_{\ell-1}(1)\psi_i \\
 -
 \frac{2\sigma}{\pi}
\sum_{j\ne i} w_j 
 \left\{
 \frac{2 p_j^2}{(p_j^2-p_i^2)^2}
 \left[ P_\ell(y_{ij}) \psi_j-\psi_i \right] 
 -  \frac{p_i\psi'_i}{p_j^2-p_i^2}
 -
\frac{Q_0(y_{ij})}{2p_i^2}
\left[ 
 P'_\ell(y_{ij})\psi_j - \frac{p_i}{p_j} 
 P'_\ell(1)\psi_i
 \right]
 +\frac{W'_{\ell-1}(y_{ij})}{2p_i^2}\psi_j
 \right\} \\
 - \frac{2\sigma}{\pi} w_i
 \left\{ \frac{3}{4p_i} \psi'_i + \frac{1}{4}  \left[  \frac{P'_\ell(1)}{p_i^2} \psi_i +\psi''_i \right] + \frac{W'_{\ell-1}(1)}{2p_i^2}\psi_i \right\}
 = E \psi_i \, .
 \label{eq:SEldisc}
\end{multline}
\end{widetext}
Note that, for both potentials, the summations over $j$ exclude the diagonal terms $j=i$, whose explicit expressions are written right after the corresponding summation.

As already mentioned above, we approximate now the derivatives $\psi'_\ell(p_i)$ and $\psi''_\ell(p_i)$ through the derivatives of an interpolating function, which represents $\psi_\ell(p)$ as a linear combination of the wave function at some subset of the grid points,
\begin{equation}
\psi_\ell(p)=\sum_j \psi_\ell(p_j) L_j(p) \, ,
\end{equation}
where the functions $L_j(p)$ are often taken to be polynomials. For instance, for Lagrange interpolation using $N_L$ points $\{ p_k \}$ near point $p$, they are
\begin{equation}
L_j(p)=\prod_{\substack{k=1 \\ k \neq j}}^{N_L} \frac{p-p_k}{p_j-p_k} \, .
\end{equation}
With this we get the derivatives at the quadrature points as 
\begin{equation}
\label{eq:Lintderiv}
\psi'_\ell(p_i)=\sum_j  L'_j(p_i)\psi_\ell(p_j)  \, , \quad  \psi''_\ell(p_i)=\sum_j L''_j(p_i) \psi_\ell(p_j)  \, .
\end{equation}
How many and exactly which indices $j$ are summed over here depends on the order of the interpolating polynomials.
It seems best to choose the interpolation points $p_k$ to be the closest quadrature points to the left and right of $p_i$. For instance, for a 5-point Lagrange interpolation, the $p_k$ should be
$p_{i-2},p_{i-1},p_i,p_{i+1},p_{i+2}$. Exceptions must be made for the points close to the ends, where not enough interpolation points to the left or right are available.

Equation (\ref{eq:SEldisc}) can now be written as an eigenvalue equation of the form $M \psi = E \psi$, where $\psi$ is a column vector with elements $\psi_i \equiv \psi_\ell(p_i)$, and $M$ is a matrix (with elements $M_{ij}$). For the derivatives we write  
\begin{equation}
\label{eq:interpol}
\psi'_i = \sum_j D^{(1)}_{ij} \psi_j \, , \qquad  \psi''_i = \sum_j D^{(2)}_{ij} \psi_j \, ,
\end{equation}
where the matrices $D^{(1)}_{ij}=L'_j(p_i)$ and $D^{(2)}_{ij}=L''_j(p_i)$ depend only on the grid points close to $p_i$. Other interpolation methods can of course be used as well, such as spline interpolation. This will lead to different matrices $D^{(1)}_{ij}$ and $D^{(2)}_{ij}$, but the general form (\ref{eq:interpol}) will stay the same.

It is convenient to slightly reorganize (\ref{eq:SEldisc}). The subtraction terms in the summations over $j$ contain $\psi_i$ and other factors that do not depend on the summation index and can be pulled out, leaving the following sums: 
\begin{widetext}

\begin{equation}
S_i^{(1)}=\sum_{j \neq i}  w_j \frac{2p_j^2}{\left(p_j^2-p_i^2\right)^2} \, , \qquad
S_i^{(2)}=\sum_{j \neq i}  \frac{w_j}{p_j^2-p_i^2}  \, , \qquad
S_i^{(3)}=\sum_{j \neq i}  \frac{w_j}{p_j} Q_0(y_{ij}) \, .
\end{equation}
Then we get
\begin{multline}
\left\{\frac{p_i^2}{2m_R} 
+\frac{\alpha p_i}{\pi}\left( S_i^{(3)} -\frac{\pi^2}{2} + \frac{w_i W_{\ell-1}(1)}{p_i} \right)
+\frac{2\sigma}{\pi}
 \left[
 S_i^{(1)}+
 \frac{P'_\ell(1)}{2p_i}\left(
 \frac{\pi^2}{2} - S_i^{(3)} - \frac{w_i}{2p_i}\right)
 \right]
\right\}
\psi_i
\\
 -\frac{\alpha}{\pi p_i }
\sum_{j \ne i} w_j \,  
  p_j\,  Q_\ell(y_{ij})\psi_j 
 -
 \frac{2\sigma}{\pi}
\sum_{j\ne i} w_j 
 \left[
 \frac{2 p_j^2}{(p_j^2-p_i^2)^2}
 P_\ell(y_{ij}) 
 -
\frac{Q_0(y_{ij})}{2p_i^2}
 P'_\ell(y_{ij}) 
 \right] \psi_j 
 \\
  + \frac{2\sigma}{\pi}\sum_j 
 \left[
 \left( p_i S_i^{(2)}-\frac{3w_i}{4p_i}\right) D_{ij}^{(1)}-
 \frac{w_i}{4}  D_{ij}^{(2)} -\frac{w_j W'_{\ell-1}(y_{ij})}{2p_i^2}
\right]
 \psi_j
 = E \psi_i \, .
 \label{eq:SEldisc2}
\end{multline}
Equation (\ref{eq:SEldisc2}) is organized in such a way that all diagonal kernel elements appear in the first line, the off-diagonal elements in the second line, and in the third line the contributions from the derivatives of the wave function, which are usually contained in a more or less narrow band along the diagonal, and a term involving $W'_{\ell-1}(y_{ij})$ that is present on and off the diagonal. 

The form of this equation is
\begin{equation}
\label{eq:SEmatrix1}
\sum_{j=1}^N M_{ij} \psi_j = E \psi_i \, ,
\end{equation}
and we can read off the elements of the matrix $M$ from (\ref{eq:SEldisc2}):
\begin{multline}
M_{ij}=
\delta_{ij}
\left\{\frac{p_i^2}{2m_R} 
+\frac{\alpha p_i}{\pi}\left( S_i^{(3)} -\frac{\pi^2}{2} + \frac{w_i W_{\ell-1}(1)}{p_i} \right)
+\frac{2\sigma}{\pi}
 \left[
S_i^{(1)}+
 \frac{P'_\ell(1)}{2p_i}\left(
 \frac{\pi^2}{2} - S_i^{(3)} - \frac{w_i}{2p_i}\right)
 \right]
\right\}
\\
-\frac{\alpha}{\pi }(1-\delta_{ij})
 \frac{w_j p_j}{p_i }
   Q_\ell(y_{ij}) 
 -
 \frac{2\sigma}{\pi}(1-\delta_{ij}) w_j
 \left[
 \frac{2 p_j^2}{(p_j^2-p_i^2)^2}
 P_\ell(y_{ij}) 
 -
\frac{Q_0(y_{ij})}{2p_i^2}
 P'_\ell(y_{ij}) 
 \right]  
 \\
  + \frac{2\sigma}{\pi} 
 \left[
 \left( p_i S_i^{(2)}-\frac{3w_i}{4p_i}\right) D_{ij}^{(1)}-
 \frac{w_i}{4}  D_{ij}^{(2)}
 -\frac{w_j W'_{\ell-1}(y_{ij})}{2p_i^2}\right]
 \, .
 \label{eq:SEmatrix}
\end{multline}

\end{widetext}

\section{Numerical results}
\label{sec:num}

\begin{table*}[b]
\caption{Energy eigenvalues for the lowest eigenstates $n=1$ to $n=10$ for the linear potential for $\ell=0$ for increasing number $N$ of Gauss-Legendre integration points, using a 5-point Lagrange interpolation. One can see an excellent convergence to the exact eigenvalues shown in the last row. The energies are given in units of $(\sigma^2/2m_R)^{1/3}$. }
\label{tab:swave-Nconv}
\begin{ruledtabular}
 \begin{tabular}{r|ccccc}
 & \multicolumn{5}{@{}c@{}}{Energy level $n$} \\
N  & 1 & 2 & 3 & 4 & 5 \\
\hline
  50 & 2.3381366586 & 4.0883811896 & 5.5205090235 & 6.7830986753 & 7.9303780220 \\
 100 & 2.3381083587 & 4.0879637759 & 5.5205602995 & 6.7865928385 & 7.9436701246 \\
 200 & 2.3381074406 & 4.0879499019 & 5.5205598624 & 6.7867044514 & 7.9441187098 \\
 400 & 2.3381074114 & 4.0879494586 & 5.5205598293 & 6.7867079757 & 7.9441331176 \\
 600 & 2.3381074106 & 4.0879494460 & 5.5205598283 & 6.7867080750 & 7.9441335251 \\
 800 & 2.3381074105 & 4.0879494446 & 5.5205598281 & 6.7867080865 & 7.9441335724 \\
 1000 & 2.3381074105 & 4.0879494443 & 5.5205598281 & 6.7867080889 & 7.9441335823 \\
 \text{Exact} & 2.3381074105 & 4.0879494441 & 5.5205598281 & 6.7867080901 & 7.9441335871 \\
\end{tabular} 
 \vspace{5mm}
\begin{tabular}{r|ccccc}
 & \multicolumn{5}{@{}c@{}}{Energy level $n$} \\
 N & 6 & 7 & 8 & 9 & 10 \\
\hline
 50 & 8.9868873730 & 9.9628553476 & 10.8600670363 & 11.6520490076 & 12.3644392700 \\
 100 & 9.0213956811 & 10.0373646777 & 11.0029378425 & 11.9257952285 & 12.8112165757 \\
 200 & 9.0226099870 & 10.0400815207 & 11.0083369103 & 11.9356672738 & 12.8281686672 \\
 400 & 9.0226495588 & 10.0401713896 & 11.0085183186 & 11.9360043877 & 12.8287571438 \\
 600 & 9.0226506823 & 10.0401739512 & 11.0085235116 & 11.9360140829 & 12.8287741529 \\
 800 & 9.0226508127 & 10.0401742488 & 11.0085241154 & 11.9360152111 & 12.8287761342 \\
 1000 & 9.0226508400 & 10.0401743111 & 11.0085242419 & 11.9360154477 & 12.8287765499 \\
 \text{Exact} & 9.0226508533 & 10.0401743416 & 11.0085243037 & 11.9360155632 & 12.8287767529 \\
\end{tabular}
\end{ruledtabular}
\label{tab:1}
\end{table*}

\begin{table*}[tb]
\caption{Energy eigenvalues for the lowest eigenstates $n=1$ to $n=10$ for the linear potential for $l=0$, calculate with different methods. The first line displays the results of Ref.~\cite{Leitao:2014jha} where the wave functions are expanded into a basis of 64 B-splines. For the results of the second line, a cubic spline interpolation with vanishing first derivatives at the endpoints was employed instead of Lagrange interpolation. The results otained with different orders of the Lagrange polynomials in lines 3--9 were calculated with 1000 integration points, the last line is the exact result of Eq.~(\ref{eq:EAiry}). The energies are given in units of $(\sigma^2/2m_R)^{1/3}$. }\label{tab:methods}%
\begin{ruledtabular}
 \begin{tabular}{l|ccccc}
 & \multicolumn{5}{@{}c@{}}{Energy level $n$} \\
Method  & 1 & 2 & 3 & 4 & 5 \\
\hline
 Expansion into B-splines~\cite{Leitao:2014jha} & 2.3381076139 & 4.0879494012 & 5.5205596162 & 6.7867079447 & 7.9441334370 \\
 Cubic spline interpolation & 2.3381073948 & 4.0879490817 & 5.5205592518 & 6.7867079666 & 7.9441352618  \\
 \text{3-point Lagrange interpolation} & 2.3381074769 & 4.0879489796 &
   5.5205581177 & 6.7867039490 & 7.9441251616 \\
 \text{4-point Lagrange interpolation} & 2.3381074265 & 4.0879498190 &
   5.5205604254 & 6.7867082210 & 7.9441318589 \\
 \text{5-point Lagrange interpolation} & 2.3381074105 & 4.0879494443 &
   5.5205598281 & 6.7867080889 & 7.9441335823 \\
 \text{6-point Lagrange interpolation} & 2.3381074105 & 4.0879494441 &
   5.5205598284 & 6.7867080913 & 7.9441335902 \\
 \text{7-point Lagrange interpolation} & 2.3381074105 & 4.0879494441 &
   5.5205598281 & 6.7867080901 & 7.9441335871 \\
 \text{8-point Lagrange interpolation} & 2.3381074105 & 4.0879494441 &
   5.5205598281 & 6.7867080901 & 7.9441335871 \\
 \text{9-point Lagrange interpolation} & 2.3381074105 & 4.0879494441 &
   5.5205598281 & 6.7867080901 & 7.9441335871 \\ \text{Exact} & 2.3381074105 & 4.0879494441 & 5.5205598281 & 6.7867080901 & 7.9441335871 \\
\end{tabular} 
 \vspace{5mm}
\begin{tabular}{l|ccccc}
 & \multicolumn{5}{@{}c@{}}{Energy level $n$} \\
 Method & 6 & 7 & 8 & 9 & 10 \\
\hline
  Expansion into B-splines~\cite{Leitao:2014jha}  & 9.0226512090 &  10.0401766855 &  11.0085333135 &  11.9360442615 &  12.8288594552 \\
 Cubic spline interpolation & 9.0226565188 & 10.0401871972
   & 11.0085487188 & 11.9360572362 & 12.8288428719 \\
 \text{3-point Lagrange interpolation} & 9.0226354555 & 10.0401482812 &
   11.0084827276 & 11.9359522932 & 12.8286841273 \\
 \text{4-point Lagrange interpolation} & 9.0226449940 & 10.0401610335 &
   11.0084990145 & 11.9359723795 & 12.8287082138 \\
 \text{5-point Lagrange interpolation} & 9.0226508400 & 10.0401743111 &
   11.0085242419 & 11.9360154477 & 12.8287765499 \\
 \text{6-point Lagrange interpolation} & 9.0226508587 & 10.0401743485 &
   11.0085243088 & 11.9360155581 & 12.8287767217 \\
 \text{7-point Lagrange interpolation} & 9.0226508533 & 10.0401743415 &
   11.0085243037 & 11.9360155630 & 12.8287767523 \\
 \text{8-point Lagrange interpolation} & 9.0226508534 & 10.0401743416 &
   11.0085243038 & 11.9360155634 & 12.8287767531 \\
 \text{9-point Lagrange interpolation} & 9.0226508533 & 10.0401743416 &
   11.0085243037 & 11.9360155632 & 12.8287767529 \\
 \text{Exact} & 9.0226508533 & 10.0401743416 & 11.0085243037 & 11.9360155632 & 12.8287767529 \\
\end{tabular}
\end{ruledtabular}
\end{table*}

\begin{table*}[tb]
\caption{Energy eigenvalues calculated with a grid of $N=200$ integration points for the lowest eigenstates $n=1$ to $n=10$ for the linear potential for $\ell=0$ for increasing number $N_L$ of points used in the Lagrange interpolation. The second column shows the exact values obtained from (\ref{eq:EAiry}), the other columns show the differences between the numerical and the exact results, $E_\mathrm{num}-E_\mathrm{exact}$, where the exponents of base 10 are given in parentheses. The energies are given in units of $(\sigma^2/2m_R)^{1/3}$. }\label{tab:N=200}%
\begin{ruledtabular}
 \begin{tabular}{ccccccccc}
 $n$ & \text{$\ell=0$ exact} & \text{$N_L$=3} & \text{$N_L$=5} & \text{$N_L$=7} & \text{$N_L$=9} & \text{$N_L$=11} & \text{$N_L$=13}
   & \text{$N_L$=15} \\
\hline
 1 & 2.33810741045976703849 & \text{8.25 (-6)} & \text{3.01 (-8)} & \text{-2.78 (-10)} & \text{1.07 (-12)} &
   \text{-1.67 (-13)} & \text{-3.23 (-7)} & \text{1.34 (-6)} \\
 2 & 4.08794944413097061664 & \text{-5.77 (-5)} & \text{4.58 (-7)} & \text{2.22 (-9)} & \text{-9.61 (-11)} &
   \text{8.98 (-13)} & \text{-3.24 (-7)} & \text{1.34 (-6)} \\
 3 & 5.52055982809555105913 & \text{-2.12 (-4)} & \text{3.43 (-8)} & \text{4.50 (-8)} & \text{-2.75 (-10)} &
   \text{-2.77 (-11)} & \text{-3.25 (-7)} & \text{1.34 (-6)} \\
 4 & 6.78670809007175899878 & \text{-5.14 (-4)} & \text{-3.64 (-6)} & \text{1.52 (-7)} & \text{4.28 (-9)} &
   \text{-2.43 (-10)} & \text{-3.25 (-7)} & \text{1.34 (-6)} \\
 5 & 7.94413358712085312314 & \text{-1.04 (-3)} & \text{-1.49 (-5)} & \text{2.43 (-7)} & \text{2.98 (-8)} &
   \text{-3.10 (-10)} & \text{-3.26 (-7)} & \text{1.34 (-6)} \\
 6 & 9.02265085334098038016 & \text{-1.91 (-3)} & \text{-4.09 (-5)} & \text{-3.61 (-8)} & \text{1.06 (-7)} &
   \text{3.45 (-9)} & \text{-3.27 (-7)} & \text{1.34 (-6)} \\
 7 & 10.04017434155808593059 & \text{-3.22 (-3)} & \text{-9.28 (-5)} & \text{-1.57 (-6)} & \text{2.60 (-7)} &
   \text{2.47 (-8)} & \text{-3.28 (-7)} & \text{1.34 (-6)} \\
 8 & 11.00852430373326289324 & \text{-5.14 (-3)} & \text{-1.87 (-4)} & \text{-6.19 (-6)} & \text{4.67 (-7)} &
   \text{9.70 (-8)} & \text{-3.26 (-7)} & \text{1.34 (-6)} \\
 9 & 11.93601556323626251701 & \text{-7.81 (-3)} & \text{-3.48 (-4)} & \text{-1.72 (-5)} & \text{5.29 (-7)} &
   \text{2.84 (-7)} & \text{-3.06 (-7)} & \text{1.34 (-6)} \\
 10 & 12.82877675286575720041 & \text{-1.14 (-2)} & \text{-6.08 (-4)} & \text{-4.02 (-5)} & \text{-1.37 (-7)} &
   \text{6.73 (-7)} & \text{-2.27 (-7)} & \text{1.33 (-6)} \\
\end{tabular}
\end{ruledtabular}
\label{tab:23new}
\end{table*}

\begin{table*}[tb]
\caption{Energy eigenvalues calculated with a grid of $N=600$ integration points for the lowest eigenstates $n=1$ to $n=10$ for the linear potential for $\ell=0$ for increasing number $N_L$ of points used in the Lagrange interpolation. The second column shows the exact values obtained from (\ref{eq:EAiry}), the other columns show the differences between the numerical and the exact results, where the exponents of base 10 are given in parentheses. The energies are given in units of $(\sigma^2/2m_R)^{1/3}$. }\label{tab:N=600}%
\begin{ruledtabular}
 \begin{tabular}{ccccccccc}
 $n$ & \text{$\ell=0$ exact} & \text{$N_L$=3} & \text{$N_L$=5} & \text{$N_L$=7} & \text{$N_L$=9} & \text{$N_L$=11} & \text{$N_L$=13}
   & \text{$N_L$=15} \\
\hline
 1 & 2.33810741045976703849 & \text{3.07 (-7)} & \text{1.25 (-10)} & \text{-1.27 (-13)} & \text{1.04 (-16)} &
   \text{2.60 (-15)} & \text{1.01 (-15)} & \text{-5.31 (-11)} \\
 2 & 4.08794944413097061664 & \text{-2.15 (-6)} & \text{1.91 (-9)} & \text{1.03 (-12)} & \text{-5.82 (-15)} &
   \text{1.32 (-15)} & \text{-2.94 (-16)} & \text{-5.31 (-11)} \\
 3 & 5.52055982809555105913 & \text{-7.91 (-6)} & \text{1.67 (-10)} & \text{2.10 (-11)} & \text{-1.57 (-14)} &
   \text{8.95 (-16)} & \text{-2.61 (-16)} & \text{-5.31 (-11)} \\
 4 & 6.78670809007175899878 & \text{-1.91 (-5)} & \text{-1.51 (-8)} & \text{7.16 (-11)} & \text{2.22 (-13)} &
   \text{6.87 (-16)} & \text{-9.53 (-16)} & \text{-5.31 (-11)} \\
 5 & 7.94413358712085312314 & \text{-3.90 (-5)} & \text{-6.20 (-8)} & \text{1.17 (-10)} & \text{1.57 (-12)} &
   \text{-6.17 (-16)} & \text{-9.39 (-16)} & \text{-5.31 (-11)} \\
 6 & 9.02265085334098038016 & \text{-7.12 (-5)} & \text{-1.71 (-7)} & \text{-4.35 (-12)} & \text{5.70 (-12)} &
   \text{2.06 (-14)} & \text{-1.17 (-15)} & \text{-5.31 (-11)} \\
 7 & 10.04017434155808593059 & \text{-1.20 (-4)} & \text{-3.90 (-7)} & \text{-7.12 (-10)} & \text{1.44 (-11)} &
   \text{1.48 (-13)} & \text{-1.34 (-15)} & \text{-5.31 (-11)} \\
 8 & 11.00852430373326289324 & \text{-1.92 (-4)} & \text{-7.92 (-7)} & \text{-2.88 (-9)} & \text{2.72 (-11)} &
   \text{6.03 (-13)} & \text{-2.15 (-16)} & \text{-5.31 (-11)} \\
 9 & 11.93601556323626251701 & \text{-2.92 (-4)} & \text{-1.48 (-6)} & \text{-8.15 (-9)} & \text{3.49 (-11)} &
   \text{1.83 (-12)} & \text{1.23 (-14)} & \text{-5.31 (-11)} \\
 10 & 12.82877675286575720041 & \text{-4.28 (-4)} & \text{-2.60 (-6)} & \text{-1.93 (-8)} & \text{7.66 (-12)} &
   \text{4.58 (-12)} & \text{7.01 (-14)} & \text{-5.31 (-11)} \\
\end{tabular}
\end{ruledtabular}
\label{tab:3new}
\end{table*}
%

\begin{table*}[tb!]
\caption{Energy eigenvalues calculated with a grid of $N=1000$ integration points for the lowest eigenstates $n=1$ to $n=10$ for the linear potential for $\ell=0$ for increasing number $N_L$ of points used in the Lagrange interpolation. The second column shows the exact values obtained from (\ref{eq:EAiry}), the other columns show the differences between the numerical and the exact results, where the exponents of base 10 are given in parentheses. The energies are given in units of $(\sigma^2/2m_R)^{1/3}$. }\label{tab:N=1000}%
\begin{ruledtabular}
 \begin{tabular}{ccccccccc}
 $n$ & \text{$\ell=0$ exact} & \text{$N_L$=3} & \text{$N_L$=5} & \text{$N_L$=7} & \text{$N_L$=9} & \text{$N_L$=11} & \text{$N_L$=13}
   & \text{$N_L$=15} \\
\hline
 1 & 2.33810741045976703849 & \text{6.64 (-8)} & \text{9.74 (-12)} & \text{-5.07 (-15)} & \text{2.37 (-15)} &
   \text{1.09 (-15)} & \text{1.23 (-15)} & \text{-1.72 (-16)} \\
 2 & 4.08794944413097061664 & \text{-4.64 (-7)} & \text{1.48 (-10)} & \text{2.83 (-14)} & \text{2.22 (-15)} &
   \text{8.83 (-16)} & \text{1.31 (-16)} & \text{-2.25 (-15)} \\
 3 & 5.52055982809555105913 & \text{-1.71 (-6)} & \text{1.32 (-11)} & \text{5.89 (-13)} & \text{2.00 (-16)} &
   \text{1.46 (-15)} & \text{-4.58 (-16)} & \text{-1.14 (-15)} \\
 4 & 6.78670809007175899878 & \text{-4.14 (-6)} & \text{-1.18 (-9)} & \text{2.01 (-12)} & \text{2.69 (-15)} &
   \text{-7.20 (-16)} & \text{-3.19 (-17)} & \text{-2.31 (-15)} \\
 5 & 7.94413358712085312314 & \text{-8.43 (-6)} & \text{-4.83 (-9)} & \text{3.30 (-12)} & \text{1.64 (-14)} &
   \text{2.17 (-16)} & \text{4.44 (-16)} & \text{-2.96 (-15)} \\
 6 & 9.02265085334098038016 & \text{-1.54 (-5)} & \text{-1.33 (-8)} & \text{-9.48 (-14)} & \text{5.74 (-14)} &
   \text{-7.26 (-16)} & \text{1.39 (-16)} & \text{-2.63 (-15)} \\
 7 & 10.04017434155808593059 & \text{-2.61 (-5)} & \text{-3.04 (-8)} & \text{-2.00 (-11)} & \text{1.48 (-13)} &
   \text{4.55 (-17)} & \text{-2.95 (-17)} & \text{-1.99 (-15)} \\
 8 & 11.00852430373326289324 & \text{-4.16 (-5)} & \text{-6.18 (-8)} & \text{-8.10 (-11)} & \text{2.79 (-13)} &
   \text{2.64 (-15)} & \text{8.66 (-16)} & \text{-8.48 (-16)} \\
 9 & 11.93601556323626251701 & \text{-6.33 (-5)} & \text{-1.16 (-7)} & \text{-2.29 (-10)} & \text{3.61 (-13)} &
   \text{7.41 (-15)} & \text{2.40 (-16)} & \text{-1.18 (-15)} \\
 10 & 12.82877675286575720041 & \text{-9.26 (-5)} & \text{-2.03 (-7)} & \text{-5.44 (-10)} & \text{9.31 (-14)} &
   \text{1.81 (-14)} & \text{1.80 (-15)} & \text{-1.25 (-15)} \\
\end{tabular}
\end{ruledtabular}
\label{tab:2new}
\end{table*}

\begin{table*}[tb!]
\caption{Energy eigenvalues for the lowest eigenstates $n=1$ to $n=10$ of the linear potential for increasing orbital angular momentum $\ell$. To be sure that all shown 10 decimal places are converged, the results were calculated with $N=1000$ integration points and a 15-point Lagrange interpolation. The energies are given in units of $(\sigma^2/2m_R)^{1/3}$. }
\label{tab:orbital}
\begin{ruledtabular}
 \begin{tabular}{c|cccccc}
 & \multicolumn{6}{@{}c@{}}{Orbital angular momentum $\ell$} \\
 $n$ & 0 & 1 & 2 & 3 & 4 & 5 \\
\hline
 1 & 2.3381074105 & 3.3612545287 & 4.2481822704 & 5.0509256508 & 5.7944228052 & 6.4930175133 \\
 2 & 4.0879494441 & 4.8844518636 & 5.6297084371 & 6.3321154965 & 6.9992605338 & 7.6369670555 \\
 3 & 5.5205598281 & 6.2076233326 & 6.8688828130 & 7.5046459023 & 8.1172598797 & 8.7092259071 \\
 4 & 6.7867080901 & 7.4056654987 & 8.0097031246 & 8.5971174402 & 9.1683269015 & 9.7243497970 \\
 5 & 7.9441335871 & 8.5152343949 & 9.0770033453 & 9.6272676380 & 10.1656642600 & 10.6924917742 \\
 6 & 9.0226508533 & 9.5576160388 & 10.0864602026 & 10.6070045389 & 11.1185630862 & 11.6210674240 \\
 7 & 10.0401743416 & 10.5465223126 & 11.0487412666 & 11.5447896633 & 12.0338590400 & 12.5156900526 \\
 8 & 11.0085243037 & 11.4914275107 & 11.9715080333 & 12.4468977833 & 12.9167581611 & 13.3807317800 \\
 9 & 11.9360155632 & 12.3992183084 & 12.8604916131 & 13.3181394690 & 13.7713365311 & 14.2196783310 \\
 10 & 12.8287767529 & 13.2750964716 & 13.7201176498 & 14.1623023780 & 14.6008585064 & 15.0353631723 \\
\end{tabular}
\end{ruledtabular}

\end{table*}

In this section we present the results of calculations we performed for the linear confining part of the Cornell potential with the Nystr\"om method, by numerically determining the eigenvalues and eigenvectors of the matrix $M$ of (\ref{eq:SEmatrix}), where we have set $\alpha=0$. 

The reason for this restriction is simply that the Coulomb problem has already been solved with the Nystr\"om method in the past~\cite{Kwon:1978,Norbury:1994bh}. We have of course also solved the pure Coulomb problem as a test case, but we do not display here any results because there are no new insights to be gained. Suffice it to say that we reproduce the exactly known energy levels and wave functions with excellent accuracy.

For the quadrature rule (\ref{eq:Quad}), we start
with Gauss-Legendre quadrature points $\{ x_i\}$ and weights $\{ w_{x,i}\}$ for the interval $x \in [-1,+1]$, such that
\begin{equation}
\int_{-1}^{+1} g(x) dx \approx \sum_{i=1}^N w_{x,i} \, g(x_i) \, ,
\end{equation}
and then map from the interval $[-1,+1]$ to $[0,\infty)$ with the transformation 
\begin{equation}
\label{eq:pmap}
p(x)=p_0 \frac{1+x}{1-x}  \, ,
\end{equation}
such that 
\begin{equation}
p_i = p_0 \frac{1+x_i}{1-x_i} \,  , \qquad w_i =  p_0 \frac{2 w_{x,i}}{(1-x_i)^2} \, ,
\end{equation}
where $p_0>0$ is a scaling parameter.

The Hamiltonian contains two parameters, namely the string tension $\sigma$ of the linear potential and the reduced mass $m_R$ of the 2-body system. Because the energy eigenvalues
scale with $(\sigma^2/2m_R)^{1/3}$, it is sufficient to solve Eq.~(\ref{eq:SEmatrix1}) for $\sigma = 2m_R = 1$.

The linear potential in S-waves is the ideal case to test our numerical methods, because its exact solutions in coordinate space are known in terms of the Airy functions ${\rm Ai}$, with the corresponding energies given by
\begin{equation}
\label{eq:EAiry}
 E_n^{\ell=0}=-z_n\left(\frac {\sigma^2}{2m_R}\right)^{1/3}\,,\, \text{with}\quad {\rm Ai} (z_n)=0\,,
 \end{equation}
where $z_n<0$ is the $n$th root of the Airy function ${\rm Ai}(z)$.
 
First we demonstrate for the linear potential the numerical convergence of the energy eigenvalues to the exact values as the number of Gauss-Legendre points $N$ increases. We calculated the lowest energy eigenvalues $n=1$ to $n=10$ using 5-point Lagrange interpolation (i.e., fourth-order polynomials) to approximate the first and second derivatives of the wave function. Table~\ref{tab:swave-Nconv} shows that our numerical S-wave energies converge quickly to the exact eigenvalues. For $N=800$ Gauss-Legendre points we already find perfect agreement in all ten displayed decimal places for the lowest eigenvalue, but also for the 10th eigenvalue we still reach very good agreement, with first deviations showing up only in the 7th decimal place.

In Table \ref{tab:methods} the S-wave results for the energy eigenvalues $n=1$ to $n=10$, obtained with four different methods, are compared. For the Nystr\"om methods, 1000 Gauss-Legendre integration points were used, while for the Galerkin method of Ref.~\cite{Leitao:2014jha}, with an expansion in a basis of 64 B-spline functions, the numerical integration was performed using the adaptive ``NIntegrate'' routine in \textit{Mathematica}. Excellent agreement with the exact results was achieved with 5-point and higher Lagrange interpolation, whereas the B-spline method with 64 basis functions still yields accurate results, but clearly less than what is achieved with the Nystr\"om methods.    

Note that the cubic spline and 4-point Lagrange interpolations both involve cubic polynomials, but they are not identical. Neighboring splines are linked through continuity conditions, such that their corresponding matrices $D^{(1)}_{ij}$ and $D^{(2)}_{ij}$ of (\ref{eq:interpol}) connect all points, whereas our Lagrange polynomials are local. As can be seen in Table \ref{tab:methods}, both yield results of comparable accuracy, where interestingly the exact results always lies between them.

The higher-order Lagrange interpolation results turned out to be so precise that we wanted to see how high their accuracy can go when the number of integration points and the order of the interpolating polynomials is increased. In Tables \ref{tab:N=200}, \ref{tab:N=600}, and \ref{tab:N=1000}, we show the differences between the numerical results and the exact energies given by Eq.~(\ref{eq:EAiry}) with higher precision. The exact energies are given up to 20 decimal places, together with the difference between the numerical and the exact results.

In Tab.~\ref{tab:N=200} we see that, for $N=200$ integration points, the accuracy increases with the interpolation order up to about 11-point interpolation, then it drops again to levels similar to 3-point interpolation, except that the accuracy of the higher states remains somewhat higher. Similarly, Tab.~\ref{tab:N=600} for $N=600$ shows that higher-order interpolations improve the accuracy further until a drop that happens at a higher order. For $N=1000$ in  Tab.~\ref{tab:N=1000}, a clear drop is no longer discernible, and an impressive accuracy is achieved even in the highly excited states.

We can conclude that, as expected, increasing the number of integration points improves the accuracy of the results. So does increasing the order of the Lagrange interpolation polynomial. However, if the number of integration points is low one should not choose a high-order interpolation. A plausible explanation for this behaviour is that higher-order interpolating polynomials tend to oscillate between the node points, such that the derivatives become unreliable. Increasing the number of nodes (here the integration points) reduces their relative distance and lowers this tendency, such that higher-order interpolation actually improves the results.

It should be mentioned that the calculations for Tables \ref{tab:N=200} to \ref{tab:N=1000} were performed with \textit{Mathematica} with 30-digit precision. In normal machine-precision (16 digits), the roundoff errors accumulate too much to allow such high accuracy. In applications to actual physical problems, like calculations of the meson spectra with the Cornell potential, a much lower accuracy is of course sufficient.

For higher partial waves, with $\ell > 0$, no exact solutions are available. However, given the extreme accuracy of our Nystr\"om method, we list in, Tab.~\ref{tab:orbital}, the converged energies, up to 10 decimal places, for angular momenta up to $\ell=5$, calculated with $N=1000$ and 15-point Lagrange interpolation.
Partial waves like P-, D-, and F-waves with $\ell=1, 2, 3$, respectively, are of practical importance, as they contribute to the wave functions of scalar, vector, axial-vector, and tensor mesons. Tables of the numerical values of our wave functions up to $\ell=5$ can be found in  \cite{stadler_2024_14217822}. A \textit{Mathematica} notebook with the routines used to produce the results presented in this work is available from the authors upon request. 

To give an idea about the necessary computing power, Table \ref{tab:CPU} displays typical computation times required to solve Eq.~(\ref{eq:SEldisc2}) with \textit{Mathematica} 14 on an Apple Macbook Pro laptop computer with an M1 chip. In these examples, normal machine precision was employed, because this is the most relevant case for practical applications. It can be seen that the computation time scales very closely with the square of the number of integration points. Most of the time is spent for calculating the matrix elements (\ref{eq:SEmatrix}), whereas the time for solving the eigenvalue problem is almost negligible. 

For the purposes of this work, solving the numerical problem in \textit{Mathematica} was clearly sufficient. However, from our experience with the calculations of Ref.~\cite{Leitao:2014jha} we can say that several orders of magnitude in speed can be gained by coding the problem in a compiled programming language like \textsc{Fortran}. It can then be further increased by parallelization. This can become relevant when the equations have to be solved many times, for instance for parameter fitting.

\begin{table}[tb!]
\caption{Computation times (in seconds) required for the numerical solution of Eq.~(\ref{eq:SEldisc2}) with $\alpha=0$, using \textit{Mathematica} 14 with normal machine precision. The calculations were performed on an Apple Macbook Pro with an M1 chip. For a fixed orbital angular momentum $\ell$, the computation timescales very closely with the square of the number of integration points $N$. It is almost independent of the order of the Lagrange interpolation. }
\label{tab:CPU}
\begin{ruledtabular}
 \begin{tabular}{r|rrrrrr}
 & \multicolumn{6}{@{}c@{}}{Orbital angular momentum $\ell$} \\
 $N$ & 0 & 1 & 2 & 3 & 4 & 5 \\
\hline
  50 & 0.08 & 0.07 & 0.26 & 0.64 & 1.12 & 1.72 \\
 100 & 0.26 & 0.28 & 1.03 & 2.55 & 4.49 & 6.90 \\
 200 & 1.04 & 1.13 & 4.10 & 10.19 & 18.00 & 27.55 \\
 400 & 4.28 & 4.53 & 16.54 & 40.97 & 72.18 & 110.92 \\
 600 & 9.71 & 10.31 & 37.27 & 92.04 & 162.61 & 248.75 \\
 800 & 17.18 & 18.34 & 66.28 & 164.55 & 287.96 & 440.95 \\
 1000 & 26.69 & 28.39 & 103.32 & 255.73 & 450.50 & 689.97 \\
\end{tabular}
\end{ruledtabular}

\end{table}

\section{Summary and Conclusions}\label{sec:concl}

In this work, we propose an accurate and simple method for solving the Schr\"odinger equation with a color Coulomb plus a linear confining potential in momentum space. The linear potential gives rise to a logarithmic singularity, which can be removed in the same way as the one from the Coulomb potential, but also to a Cauchy principal-value singularity, which stands in the way of a direct application of the standard Nystr\"om method for its numerical solution. 
In a preceding paper~\cite{Leitao:2014jha}, we have shown how the principal-value singularity can be subtracted, thus making the Schr\"odinger equation singularity free. However, as a consequence of this subtraction, the derivatives of the---\emph{a priori}---unknown wave function appear in the equation. 

One way of dealing with this obstacle is to use a Galerkin method, i.e., to expand the wave function in a basis of appropriate functions whose derivatives are known or can be calculated. 
In Ref.~\cite{Leitao:2014jha} we followed this approach and solved the singularity-free Schr\"odinger equation by means of an expansion in a basis of cubic B-splines. We found that the solutions converge well with increasing number of basis functions, especially for the lowest energy eigenvalues. For the higher-energy eigenstates, however, a rather large number of basis functions was needed to find good agreement with the exactly known values. 

A drawback of the expansion method is that the basis functions must be chosen and adapted for each problem very carefully. If they do not have the same asymptotic behavior as the correct solution, problems with convergence and accuracy are likely to occur. 

In this work, we have demonstrated that it is in fact possible to use the Nystr\"om method to solve the momentum-space Schr\"odinger equation with the linear confining potential, and consequently with the whole Cornell potential, when its principal-value singularity is treated with the subtraction technique of~\cite{Leitao:2014jha}. The first and second derivatives of the wave function that enter the equation can be expressed in terms of the values of the wave function at neighboring points by means of interpolating functions. The problem then reduces to an ordinary matrix eigenvalue problem that can be easily solved.

We have tried different interpolation functions, most of them Lagrange interpolating polynomials of different order, but for comparison also cubic splines. Because the solutions for the linear potential in S-waves are known analytically, the accuracy of our method with respect to the number of Gauss-Legendre quadrature points and the order of the interpolating polynomial can be rigorously evaluated.

We found that already with a relatively small number of about 100 to 200 quadrature points and a low-order Lagrange interpolation, the achieved accuracy is more than sufficient for most practical applications in quark-antiquark bound state calculations. By further increasing the number of integration points and the order of the Lagrange interpolation, extreme accuracy can be achieved. With 1000 integration points and a Lagrange interpolation of 11 or more points, the differences between the numerical and the exact results are of the order of $10^{-15}$ or smaller. The computing time is mainly determined by the number of quadrature points, whereas increasing the order of the interpolation polynomial comes at almost no additional cost. In our calculations, the Nystr\"om method was substantially faster than the B-spline expansion we used in ~\cite{Leitao:2014jha} to achieve about the same accuracy.

For partial waves other than S-waves, no exact analytic solutions are known, and therefore the accuracy of the numerical solutions cannot be tested with the same level of rigor. Nevertheless, for reference, we also provide our converged energies, up to 10 decimal places, for angular momenta up to $\ell=5$, calculated with $N=1000$ and 15-point Lagrange interpolation. 

We conclude that the Nystr\"om method we presented in this work to solve the Cornell potential in momentum space is simple, efficient, and very accurate. It requires no particular adaptation of quadrature points or weights. It is also flexible enough to be used with other interaction kernels that have a similar singularity structure, such as the relativistic corrections to the Cornell potential. In particular, we will apply this method to the solution of the CST equations for hadrons with fully relativistic kernels.

\begin{acknowledgments}
This work was supported by FCT under project No.\ CERN/FIS-PAR/0023/2021. E.B. was also supported by FCT and IST-ID under contract No. IST-ID/148/2018.  
\end{acknowledgments}

\section*{DATA
 AVAILABILITY}
The
 data that support the findings of this article are openly available~\cite{stadler_2024_14217822}.
 \appendix

\section{Approximate derivatives with Lagrange interpolation}
In this appendix, we list the explicit expressions for the matrices that determine the approximate first and second derivatives of the wave function at the grid points.
\subsection{Construction of the Lagrange polynomial}

We start with a set of $N_L$ points $\{p_j\}$, where  $j=1,\dots,N_L$, and the respective values of a function $f(p)$ at these points, $\{ f_j \}$, where we abbreviate $f_j \equiv f(p_j)$. 
The Lagrange interpolating polynomial is the unique polynomial $L(p)$  of order $N_L-1$ that passes exactly through these given points $(p_j,f_j)$, i.e.,
\begin{equation}
L(p_j)=f(p_j)\, , \quad j=1,\dots,N_L \, .
\end{equation}
It can be written in the form
\begin{equation}
\label{eq:L}
L(p)=\sum_{j=1}^{N_L} f_j L_j(p) \, ,
\end{equation}
with
\begin{equation}
L_j(p) = \prod_{\substack{k=1 \\ k\ne j}}^{N_L} \frac{p-p_k}{p_j-p_k}\,.
\end{equation}
To find the first derivative of the interpolating polynomial,
\begin{equation}
\label{eq:dL}
L'(p)=\sum_{j=1}^{N_L} f_j L'_j(p) \, ,
\end{equation}
 we only need to find $L'_j(p)$. Applying the rule for the derivative of a product, we get
\begin{equation}
\label{eq:dLj}
L'_j(p)= \sum_{\substack{m =1 \\m\ne j}}^{N_L} \frac{1}{p_j-p_m} \prod_{\substack{k=1\\k\ne j \\ k \ne m}}^{N_L} \frac{p-p_k}{p_j-p_k} \, .
\end{equation}
Similarly, we get the second derivative through
\begin{equation}
\label{eq:ddL}
L''(p)=\sum_{j=1}^{N_L} f_j L''_j(p) \, ,
\end{equation}
and
\begin{equation}
\label{eq:ddLj}
L''_j(p)= \sum_{\substack{m =1 \\m\ne j}}^{N_L} \frac{1}{p_j-p_m} \sum_{\substack{l =1 \\l\ne j \\l\ne m}}^{N_L} \frac{1}{p_j-p_l} \prod_{\substack{k=1\\k\ne j \\ k \ne l \\ k \ne m }}^{N_L} \frac{p-p_k}{p_j-p_k} \, .
\end{equation}

\subsection{Derivatives at the nodes}
An important application of the Lagrange polynomials is the numerical approximation of the derivatives of the unknown function $f(p)$ that is only known at a list of points. The derivatives at arbitrary points can be calculated from the expressions (\ref{eq:dL}) and (\ref{eq:dLj}), and (\ref{eq:ddL}) and (\ref{eq:ddLj}), respectively. If the derivatives are needed only at the nodes themselves, this can be somewhat simplified.

First, note that for any specific node $p_i$, which is one of the set  $\{p_j\}$, we get
\begin{equation}
L_j(p_i) = \prod_{\substack{k=1 \\ k\ne j}}^{N_L} \frac{p_i-p_k}{p_j-p_k}=\delta_{ij} \, ,
\end{equation}
because if $i \ne j$, $i$ will coincide with one of the $k$'s and therefore the product is zero because it must include a factor $p_i-p_i$, whereas each factor is $1$ if $i=j$. Inserting in (\ref{eq:L}), this also implies
\begin{equation}
L(p_i)=\sum_{j=1}^{N_L} f_j L_j(p_i)=\sum_{j=1}^{N_L} f_j \delta_{ij} =f_i \, ,
\end{equation}
as it must.

Next, we calculate (\ref{eq:dLj}) for $p=p_i$ 
\begin{equation}
\label{eq:dLi}
L'(p_i)=\sum_{j=1}^{N_L} f_j L'_j(p_i) \, , 
\end{equation}
\begin{equation}
\label{eq:dLjI}
L'_j(p_i)= \sum_{\substack{m =1 \\m\ne j}}^{N_L} \frac{1}{p_j-p_m} \prod_{\substack{k=1\\k\ne j \\ k \ne m }}^{N_L} \frac{p_i-p_k}{p_j-p_k} \, ,
\end{equation}
\begin{equation}
\label{eq:ddLi}
L''(p_i)=\sum_{j=1}^{N_L} f_j L''_j(p_i) \, ,
\end{equation}
\begin{equation}
\label{eq:ddLji}
L''_j(p_i)= \sum_{\substack{m =1 \\m\ne j}}^{N_L} \frac{1}{p_j-p_m} \sum_{\substack{l =1 \\l\ne j \\l\ne m}}^{N_L} \frac{1}{p_j-p_l} \prod_{\substack{k=1\\k\ne j \\ k \ne l \\ k \ne m }}^{N_L} \frac{p_i-p_k}{p_j-p_k} \, .
\end{equation}
Note that $m=i$ and $l=i$ are not excluded from the sums, and that $k=i$ gives 0 in the product.

\subsection{Interpolation on a subset of points}
Usually we have a table of $N>N_L$ points and first need to select a subset of points on which to perform the $N_L$-point Lagrange interpolation. The result of the interpolated function and its derivatives will depend on the selected subset, and clearly different reasonable choices are possible.

We adopted an algorithm for the selection of the subset of points for the case that we want to calculate the derivatives at one of the $N$ given points $p_i$. We assume that the closest neighboring points should be selected, i.e., we need to determine the offset $s$, such that the interpolation will use all $N_L$ points between $j=s+1$ and $j=s+N_L$
\begin{align}
L(p_i) & =\sum_{j=s+1}^{s+N_L} f_j L_j(p_i)=f_i \, , \\
L'(p_i) & =\sum_{j=s+1}^{s+N_L} f_j L'_j(p_i) \, , \\
\label{eq:Lag1}
L'_j(p_i) & = \sum_{\substack{m =s+1 \\m\ne j}}^{s+N_L} \frac{1}{p_j-p_m} \prod_{\substack{k=s+1\\k\ne j \\ k \ne m }}^{s+N_L} \frac{p_i-p_k}{p_j-p_k} \, , \\
L''(p_i) & =\sum_{j=s+1}^{s+N_L} f_j L''_j(p_i) \, , \\
\label{eq:Lag2}
L''_j(p_i) & = \sum_{\substack{m =s+1 \\m\ne j}}^{s+N_L} \frac{1}{p_j-p_m} \sum_{\substack{l =s+1 \\l\ne j \\l\ne m}}^{s+N_L} \frac{1}{p_j-p_l} \prod_{\substack{k=s+1\\k\ne j \\ k \ne l \\ k \ne m }}^{s+N_L} \frac{p_i-p_k}{p_j-p_k} \, .
\end{align}
With our map (\ref{eq:pmap}), the points are distributed such that the distance between them increases as one moves to larger values. In this situation, it makes sense to take an equal number of points to the left and to the right of $i$ if $N_L$ is odd, but one more point to the left if $N_L$ is even.

Our algorithm for determining $s$ is:
\begin{itemize}
\item{For odd $N_L$}
  \begin{itemize}
  \item[(i)]{If $i< \frac{N_L+1}{2}$:} $s=0$
  \item[(ii)]{If $i>N-\frac{N_L-1}{2}$:} $s=N-N_L$
  \item[(iii)]{Else:} $s=i-\frac{N_L+1}{2}$
  \end{itemize}
\item{For even $N_L$}
  \begin{itemize}
  \item[(i)]{If $i< \frac{N_L+2}{2}$:} $s=0$
  \item[(ii)]{If $i>N-\frac{N_L-2}{2}$:} $s=N-N_L$
  \item[(iii)]{Else:} $s=i-\frac{N_L+2}{2}$.
  \end{itemize}

\end{itemize}
The numerical results presented in this work used this algorithm for the calculation of the matrices
\begin{equation}
D^{(1)}_{ij}=L'_j(p_i) \, ,\quad D^{(2)}_{ij}=L''_j(p_i) \, ,
\end{equation}
as given in (\ref{eq:Lag1}) and (\ref{eq:Lag2}).

\bibliography{PapersDB}

\providecommand{\noopsort}[1]{}\providecommand{\singleletter}[1]{#1}%
\begin{thebibliography}{40}%
\makeatletter
\providecommand \@ifxundefined [1]{%
 \@ifx{#1\undefined}
}%
\providecommand \@ifnum [1]{%
 \ifnum #1\expandafter \@firstoftwo
 \else \expandafter \@secondoftwo
 \fi
}%
\providecommand \@ifx [1]{%
 \ifx #1\expandafter \@firstoftwo
 \else \expandafter \@secondoftwo
 \fi
}%
\providecommand \natexlab [1]{#1}%
\providecommand \enquote  [1]{``#1''}%
\providecommand \bibnamefont  [1]{#1}%
\providecommand \bibfnamefont [1]{#1}%
\providecommand \citenamefont [1]{#1}%
\providecommand \href@noop [0]{\@secondoftwo}%
\providecommand \href [0]{\begingroup \@sanitize@url \@href}%
\providecommand \@href[1]{\@@startlink{#1}\@@href}%
\providecommand \@@href[1]{\endgroup#1\@@endlink}%
\providecommand \@sanitize@url [0]{\catcode `\\12\catcode `\$12\catcode
  `\&12\catcode `\#12\catcode `\^12\catcode `\_12\catcode `\%12\relax}%
\providecommand \@@startlink[1]{}%
\providecommand \@@endlink[0]{}%
\providecommand \url  [0]{\begingroup\@sanitize@url \@url }%
\providecommand \@url [1]{\endgroup\@href {#1}{\urlprefix }}%
\providecommand \urlprefix  [0]{URL }%
\providecommand \Eprint [0]{\href }%
\providecommand \doibase [0]{https://doi.org/}%
\providecommand \selectlanguage [0]{\@gobble}%
\providecommand \bibinfo  [0]{\@secondoftwo}%
\providecommand \bibfield  [0]{\@secondoftwo}%
\providecommand \translation [1]{[#1]}%
\providecommand \BibitemOpen [0]{}%
\providecommand \bibitemStop [0]{}%
\providecommand \bibitemNoStop [0]{.\EOS\space}%
\providecommand \EOS [0]{\spacefactor3000\relax}%
\providecommand \BibitemShut  [1]{\csname bibitem#1\endcsname}%
\let\auto@bib@innerbib\@empty
\bibitem [{\citenamefont {Eichten}\ \emph {et~al.}(1975)\citenamefont
  {Eichten}, \citenamefont {Gottfried}, \citenamefont {Kinoshita},
  \citenamefont {Kogut}, \citenamefont {Lane},\ and\ \citenamefont
  {Yan}}]{PhysRevLett.34.369}%
  \BibitemOpen
  \bibfield  {author} {\bibinfo {author} {\bibfnamefont {E.}~\bibnamefont
  {Eichten}}, \bibinfo {author} {\bibfnamefont {K.}~\bibnamefont {Gottfried}},
  \bibinfo {author} {\bibfnamefont {T.}~\bibnamefont {Kinoshita}}, \bibinfo
  {author} {\bibfnamefont {J.}~\bibnamefont {Kogut}}, \bibinfo {author}
  {\bibfnamefont {K.~D.}\ \bibnamefont {Lane}},\ and\ \bibinfo {author}
  {\bibfnamefont {T.~M.}\ \bibnamefont {Yan}},\ }\href
  {https://doi.org/10.1103/PhysRevLett.34.369} {\bibfield  {journal} {\bibinfo
  {journal} {Phys. Rev. Lett.}\ }\textbf {\bibinfo {volume} {34}},\ \bibinfo
  {pages} {369} (\bibinfo {year} {1975})}\BibitemShut {NoStop}%
\bibitem [{\citenamefont {Eichten}\ \emph {et~al.}(1978)\citenamefont
  {Eichten}, \citenamefont {Gottfried}, \citenamefont {Kinoshita},
  \citenamefont {Lane},\ and\ \citenamefont {Yan}}]{Eichten:1978}%
  \BibitemOpen
  \bibfield  {author} {\bibinfo {author} {\bibfnamefont {E.}~\bibnamefont
  {Eichten}}, \bibinfo {author} {\bibfnamefont {K.}~\bibnamefont {Gottfried}},
  \bibinfo {author} {\bibfnamefont {T.}~\bibnamefont {Kinoshita}}, \bibinfo
  {author} {\bibfnamefont {K.~D.}\ \bibnamefont {Lane}},\ and\ \bibinfo
  {author} {\bibfnamefont {T.~M.}\ \bibnamefont {Yan}},\ }\href
  {https://doi.org/10.1103/PhysRevD.17.3090} {\bibfield  {journal} {\bibinfo
  {journal} {Phys. Rev. D}\ }\textbf {\bibinfo {volume} {17}},\ \bibinfo
  {pages} {3090} (\bibinfo {year} {1978})}\BibitemShut {NoStop}%
\bibitem [{\citenamefont {Eichten}\ \emph {et~al.}(1980)\citenamefont
  {Eichten}, \citenamefont {Gottfried}, \citenamefont {Kinoshita},
  \citenamefont {Lane},\ and\ \citenamefont {Yan}}]{Eichten:1980}%
  \BibitemOpen
  \bibfield  {author} {\bibinfo {author} {\bibfnamefont {E.}~\bibnamefont
  {Eichten}}, \bibinfo {author} {\bibfnamefont {K.}~\bibnamefont {Gottfried}},
  \bibinfo {author} {\bibfnamefont {T.}~\bibnamefont {Kinoshita}}, \bibinfo
  {author} {\bibfnamefont {K.~D.}\ \bibnamefont {Lane}},\ and\ \bibinfo
  {author} {\bibfnamefont {T.~M.}\ \bibnamefont {Yan}},\ }\href
  {https://doi.org/10.1103/PhysRevD.21.203} {\bibfield  {journal} {\bibinfo
  {journal} {Phys. Rev. D}\ }\textbf {\bibinfo {volume} {21}},\ \bibinfo
  {pages} {203} (\bibinfo {year} {1980})}\BibitemShut {NoStop}%
\bibitem [{\citenamefont {Godfrey}\ and\ \citenamefont
  {Isgur}(1985)}]{Godfrey:1985aa}%
  \BibitemOpen
  \bibfield  {author} {\bibinfo {author} {\bibfnamefont {S.}~\bibnamefont
  {Godfrey}}\ and\ \bibinfo {author} {\bibfnamefont {N.}~\bibnamefont
  {Isgur}},\ }\href {https://doi.org/10.1103/PhysRevD.32.189} {\bibfield
  {journal} {\bibinfo  {journal} {Phys. Rev. D}\ }\textbf {\bibinfo {volume}
  {32}},\ \bibinfo {pages} {189} (\bibinfo {year} {1985})}\BibitemShut
  {NoStop}%
\bibitem [{\citenamefont {Barnes}\ \emph {et~al.}(2005)\citenamefont {Barnes},
  \citenamefont {Godfrey},\ and\ \citenamefont {Swanson}}]{PhysRevD.72.054026}%
  \BibitemOpen
  \bibfield  {author} {\bibinfo {author} {\bibfnamefont {T.}~\bibnamefont
  {Barnes}}, \bibinfo {author} {\bibfnamefont {S.}~\bibnamefont {Godfrey}},\
  and\ \bibinfo {author} {\bibfnamefont {E.~S.}\ \bibnamefont {Swanson}},\
  }\href {https://doi.org/10.1103/PhysRevD.72.054026} {\bibfield  {journal}
  {\bibinfo  {journal} {Phys. Rev. D}\ }\textbf {\bibinfo {volume} {72}},\
  \bibinfo {pages} {054026} (\bibinfo {year} {2005})}\BibitemShut {NoStop}%
\bibitem [{\citenamefont {Soni}\ \emph {et~al.}(2018)\citenamefont {Soni},
  \citenamefont {Joshi}, \citenamefont {Shah}, \citenamefont {Chauhan},\ and\
  \citenamefont {Pandya}}]{Soni:2017wvy}%
  \BibitemOpen
  \bibfield  {author} {\bibinfo {author} {\bibfnamefont {N.~R.}\ \bibnamefont
  {Soni}}, \bibinfo {author} {\bibfnamefont {B.~R.}\ \bibnamefont {Joshi}},
  \bibinfo {author} {\bibfnamefont {R.~P.}\ \bibnamefont {Shah}}, \bibinfo
  {author} {\bibfnamefont {H.~R.}\ \bibnamefont {Chauhan}},\ and\ \bibinfo
  {author} {\bibfnamefont {J.~N.}\ \bibnamefont {Pandya}},\ }\href
  {https://doi.org/10.1140/epjc/s10052-018-6068-6} {\bibfield  {journal}
  {\bibinfo  {journal} {Eur. Phys. J. C}\ }\textbf {\bibinfo {volume} {78}},\
  \bibinfo {pages} {592} (\bibinfo {year} {2018})},\ \Eprint
  {https://arxiv.org/abs/1707.07144} {arXiv:1707.07144 [hep-ph]} \BibitemShut
  {NoStop}%
\bibitem [{\citenamefont {Soni}\ \emph {et~al.}(2022)\citenamefont {Soni},
  \citenamefont {Parekh}, \citenamefont {Patel}, \citenamefont {Gadaria},\ and\
  \citenamefont {Pandya}}]{Soni:2020tji}%
  \BibitemOpen
  \bibfield  {author} {\bibinfo {author} {\bibfnamefont {N.~R.}\ \bibnamefont
  {Soni}}, \bibinfo {author} {\bibfnamefont {R.~M.}\ \bibnamefont {Parekh}},
  \bibinfo {author} {\bibfnamefont {J.~J.}\ \bibnamefont {Patel}}, \bibinfo
  {author} {\bibfnamefont {A.~N.}\ \bibnamefont {Gadaria}},\ and\ \bibinfo
  {author} {\bibfnamefont {J.~N.}\ \bibnamefont {Pandya}},\ }\href
  {https://doi.org/10.1007/s00601-022-01778-6} {\bibfield  {journal} {\bibinfo
  {journal} {Few Body Syst.}\ }\textbf {\bibinfo {volume} {63}},\ \bibinfo
  {pages} {77} (\bibinfo {year} {2022})},\ \Eprint
  {https://arxiv.org/abs/2012.00294} {arXiv:2012.00294 [hep-ph]} \BibitemShut
  {NoStop}%
\bibitem [{\citenamefont {Kang}\ and\ \citenamefont {Won}(2008)}]{Kang:2006jd}%
  \BibitemOpen
  \bibfield  {author} {\bibinfo {author} {\bibfnamefont {D.}~\bibnamefont
  {Kang}}\ and\ \bibinfo {author} {\bibfnamefont {E.}~\bibnamefont {Won}},\
  }\href {https://doi.org/10.1016/j.jcp.2007.11.028} {\bibfield  {journal}
  {\bibinfo  {journal} {J. Comput. Phys.}\ }\textbf {\bibinfo {volume} {227}},\
  \bibinfo {pages} {2970} (\bibinfo {year} {2008})},\ \Eprint
  {https://arxiv.org/abs/physics/0609176} {arXiv:physics/0609176} \BibitemShut
  {NoStop}%
\bibitem [{\citenamefont {Eichten}\ and\ \citenamefont
  {Feinberg}(1981)}]{PhysRevD.23.2724}%
  \BibitemOpen
  \bibfield  {author} {\bibinfo {author} {\bibfnamefont {E.}~\bibnamefont
  {Eichten}}\ and\ \bibinfo {author} {\bibfnamefont {F.}~\bibnamefont
  {Feinberg}},\ }\href {https://doi.org/10.1103/PhysRevD.23.2724} {\bibfield
  {journal} {\bibinfo  {journal} {Phys. Rev. D}\ }\textbf {\bibinfo {volume}
  {23}},\ \bibinfo {pages} {2724} (\bibinfo {year} {1981})}\BibitemShut
  {NoStop}%
\bibitem [{\citenamefont {Eichmann}\ \emph {et~al.}(2016)\citenamefont
  {Eichmann}, \citenamefont {Sanchis-Alepuz}, \citenamefont {Williams},
  \citenamefont {Alkofer},\ and\ \citenamefont {Fischer}}]{Eichmann:2016yit}%
  \BibitemOpen
  \bibfield  {author} {\bibinfo {author} {\bibfnamefont {G.}~\bibnamefont
  {Eichmann}}, \bibinfo {author} {\bibfnamefont {H.}~\bibnamefont
  {Sanchis-Alepuz}}, \bibinfo {author} {\bibfnamefont {R.}~\bibnamefont
  {Williams}}, \bibinfo {author} {\bibfnamefont {R.}~\bibnamefont {Alkofer}},\
  and\ \bibinfo {author} {\bibfnamefont {C.~S.}\ \bibnamefont {Fischer}},\
  }\href {https://doi.org/10.1016/j.ppnp.2016.07.001} {\bibfield  {journal}
  {\bibinfo  {journal} {Prog. Part. Nucl. Phys.}\ }\textbf {\bibinfo {volume}
  {91}},\ \bibinfo {pages} {1} (\bibinfo {year} {2016})},\ \Eprint
  {https://arxiv.org/abs/1606.09602} {arXiv:1606.09602 [hep-ph]} \BibitemShut
  {NoStop}%
\bibitem [{\citenamefont {Alkofer}\ and\ \citenamefont {von
  Smekal}(2001)}]{Alkofer:2000wg}%
  \BibitemOpen
  \bibfield  {author} {\bibinfo {author} {\bibfnamefont {R.}~\bibnamefont
  {Alkofer}}\ and\ \bibinfo {author} {\bibfnamefont {L.}~\bibnamefont {von
  Smekal}},\ }\href {https://doi.org/10.1016/S0370-1573(01)00010-2} {\bibfield
  {journal} {\bibinfo  {journal} {Phys. Rept.}\ }\textbf {\bibinfo {volume}
  {353}},\ \bibinfo {pages} {281} (\bibinfo {year} {2001})}\BibitemShut
  {NoStop}%
\bibitem [{\citenamefont {Fischer}(2006)}]{Fischer:2006ub}%
  \BibitemOpen
  \bibfield  {author} {\bibinfo {author} {\bibfnamefont {C.~S.}\ \bibnamefont
  {Fischer}},\ }\href {https://doi.org/10.1088/0954-3899/32/8/R02} {\bibfield
  {journal} {\bibinfo  {journal} {J. Phys. G}\ }\textbf {\bibinfo {volume}
  {G32}},\ \bibinfo {pages} {R253} (\bibinfo {year} {2006})}\BibitemShut
  {NoStop}%
\bibitem [{\citenamefont {Roberts}\ and\ \citenamefont
  {Williams}(1994)}]{Roberts:1994dr}%
  \BibitemOpen
  \bibfield  {author} {\bibinfo {author} {\bibfnamefont {C.~D.}\ \bibnamefont
  {Roberts}}\ and\ \bibinfo {author} {\bibfnamefont {A.~G.}\ \bibnamefont
  {Williams}},\ }\href {https://doi.org/10.1016/0146-6410(94)90049-3}
  {\bibfield  {journal} {\bibinfo  {journal} {Prog.Part.Nucl.Phys.}\ }\textbf
  {\bibinfo {volume} {33}},\ \bibinfo {pages} {477} (\bibinfo {year} {1994})},\
  \Eprint {https://arxiv.org/abs/hep-ph/9403224} {arXiv:hep-ph/9403224
  [hep-ph]} \BibitemShut {NoStop}%
\bibitem [{\citenamefont {Roberts}\ and\ \citenamefont
  {Schmidt}(2000)}]{Roberts:2000aa}%
  \BibitemOpen
  \bibfield  {author} {\bibinfo {author} {\bibfnamefont {C.~D.}\ \bibnamefont
  {Roberts}}\ and\ \bibinfo {author} {\bibfnamefont {S.~M.}\ \bibnamefont
  {Schmidt}},\ }\href@noop {} {\bibfield  {journal} {\bibinfo  {journal}
  {Prog.Part.Nucl.Phys.}\ }\textbf {\bibinfo {volume} {45}},\ \bibinfo {pages}
  {S1} (\bibinfo {year} {2000})},\ \Eprint
  {https://arxiv.org/abs/nucl-th/0005064} {arXiv:nucl-th/0005064 [nucl-th]}
  \BibitemShut {NoStop}%
\bibitem [{\citenamefont {Maris}\ and\ \citenamefont
  {Roberts}(2003)}]{Maris:2003vk}%
  \BibitemOpen
  \bibfield  {author} {\bibinfo {author} {\bibfnamefont {P.}~\bibnamefont
  {Maris}}\ and\ \bibinfo {author} {\bibfnamefont {C.~D.}\ \bibnamefont
  {Roberts}},\ }\href {https://doi.org/10.1142/S0218301303001326} {\bibfield
  {journal} {\bibinfo  {journal} {Int. J. Mod. Phys.}\ }\textbf {\bibinfo
  {volume} {E12}},\ \bibinfo {pages} {297} (\bibinfo {year}
  {2003})}\BibitemShut {NoStop}%
\bibitem [{\citenamefont {Gross}(1969)}]{Gross:1969rv}%
  \BibitemOpen
  \bibfield  {author} {\bibinfo {author} {\bibfnamefont {F.}~\bibnamefont
  {Gross}},\ }\href {https://doi.org/10.1103/PhysRev.186.1448} {\bibfield
  {journal} {\bibinfo  {journal} {Phys. Rev.}\ }\textbf {\bibinfo {volume}
  {186}},\ \bibinfo {pages} {1448} (\bibinfo {year} {1969})}\BibitemShut
  {NoStop}%
\bibitem [{\citenamefont {Biernat}\ \emph
  {et~al.}(2014{\natexlab{a}})\citenamefont {Biernat}, \citenamefont {Gross},
  \citenamefont {Pe\~na},\ and\ \citenamefont {Stadler}}]{Biernat:2014a}%
  \BibitemOpen
  \bibfield  {author} {\bibinfo {author} {\bibfnamefont {E.~P.}\ \bibnamefont
  {Biernat}}, \bibinfo {author} {\bibfnamefont {F.}~\bibnamefont {Gross}},
  \bibinfo {author} {\bibfnamefont {M.~T.}\ \bibnamefont {Pe\~na}},\ and\
  \bibinfo {author} {\bibfnamefont {A.}~\bibnamefont {Stadler}},\ }\href
  {https://doi.org/10.1103/PhysRevD.89.016005} {\bibfield  {journal} {\bibinfo
  {journal} {Phys. Rev. D}\ }\textbf {\bibinfo {volume} {89}},\ \bibinfo
  {pages} {016005} (\bibinfo {year} {2014}{\natexlab{a}})}\BibitemShut
  {NoStop}%
\bibitem [{\citenamefont {Biernat}\ \emph
  {et~al.}(2014{\natexlab{b}})\citenamefont {Biernat}, \citenamefont {Pe\~na},
  \citenamefont {Ribeiro}, \citenamefont {Stadler},\ and\ \citenamefont
  {Gross}}]{Biernat:2014xaa}%
  \BibitemOpen
  \bibfield  {author} {\bibinfo {author} {\bibfnamefont {E.~P.}\ \bibnamefont
  {Biernat}}, \bibinfo {author} {\bibfnamefont {M.~T.}\ \bibnamefont {Pe\~na}},
  \bibinfo {author} {\bibfnamefont {J.~E.}\ \bibnamefont {Ribeiro}}, \bibinfo
  {author} {\bibfnamefont {A.}~\bibnamefont {Stadler}},\ and\ \bibinfo {author}
  {\bibfnamefont {F.}~\bibnamefont {Gross}},\ }\href
  {https://doi.org/10.1103/PhysRevD.90.096008} {\bibfield  {journal} {\bibinfo
  {journal} {Phys. Rev. D}\ }\textbf {\bibinfo {volume} {90}},\ \bibinfo
  {pages} {096008} (\bibinfo {year} {2014}{\natexlab{b}})},\ \Eprint
  {https://arxiv.org/abs/1408.1625} {arXiv:1408.1625 [hep-ph]} \BibitemShut
  {NoStop}%
\bibitem [{\citenamefont {Savkli}\ and\ \citenamefont
  {Gross}(2001)}]{Savkli:1999me}%
  \BibitemOpen
  \bibfield  {author} {\bibinfo {author} {\bibfnamefont {C.}~\bibnamefont
  {Savkli}}\ and\ \bibinfo {author} {\bibfnamefont {F.}~\bibnamefont {Gross}},\
  }\href {https://doi.org/10.1103/PhysRevC.63.035208} {\bibfield  {journal}
  {\bibinfo  {journal} {Phys. Rev. C}\ }\textbf {\bibinfo {volume} {63}},\
  \bibinfo {pages} {035208} (\bibinfo {year} {2001})},\ \Eprint
  {https://arxiv.org/abs/hep-ph/9911319} {arXiv:hep-ph/9911319 [hep-ph]}
  \BibitemShut {NoStop}%
\bibitem [{\citenamefont {Gross}\ and\ \citenamefont
  {Milana}(1991)}]{Gross:1991te}%
  \BibitemOpen
  \bibfield  {author} {\bibinfo {author} {\bibfnamefont {F.}~\bibnamefont
  {Gross}}\ and\ \bibinfo {author} {\bibfnamefont {J.}~\bibnamefont {Milana}},\
  }\href {https://doi.org/10.1103/PhysRevD.43.2401} {\bibfield  {journal}
  {\bibinfo  {journal} {Phys. Rev. D}\ }\textbf {\bibinfo {volume} {43}},\
  \bibinfo {pages} {2401} (\bibinfo {year} {1991})}\BibitemShut {NoStop}%
\bibitem [{\citenamefont {Gross}\ and\ \citenamefont
  {Milana}(1992)}]{Gross:1991pk}%
  \BibitemOpen
  \bibfield  {author} {\bibinfo {author} {\bibfnamefont {F.}~\bibnamefont
  {Gross}}\ and\ \bibinfo {author} {\bibfnamefont {J.}~\bibnamefont {Milana}},\
  }\href {https://doi.org/10.1103/PhysRevD.45.969} {\bibfield  {journal}
  {\bibinfo  {journal} {Phys. Rev. D}\ }\textbf {\bibinfo {volume} {45}},\
  \bibinfo {pages} {969} (\bibinfo {year} {1992})}\BibitemShut {NoStop}%
\bibitem [{\citenamefont {Gross}\ and\ \citenamefont
  {Milana}(1994)}]{Gross:1994he}%
  \BibitemOpen
  \bibfield  {author} {\bibinfo {author} {\bibfnamefont {F.}~\bibnamefont
  {Gross}}\ and\ \bibinfo {author} {\bibfnamefont {J.}~\bibnamefont {Milana}},\
  }\href {https://doi.org/10.1103/PhysRevD.50.3332} {\bibfield  {journal}
  {\bibinfo  {journal} {Phys. Rev. D}\ }\textbf {\bibinfo {volume} {50}},\
  \bibinfo {pages} {3332} (\bibinfo {year} {1994})}\BibitemShut {NoStop}%
\bibitem [{\citenamefont {Gross}(1964{\natexlab{a}})}]{Gross:1964mla}%
  \BibitemOpen
  \bibfield  {author} {\bibinfo {author} {\bibfnamefont {F.}~\bibnamefont
  {Gross}},\ }\href {https://doi.org/10.1103/PhysRev.134.B405} {\bibfield
  {journal} {\bibinfo  {journal} {Phys. Rev.}\ }\textbf {\bibinfo {volume}
  {134}},\ \bibinfo {pages} {B405} (\bibinfo {year}
  {1964}{\natexlab{a}})}\BibitemShut {NoStop}%
\bibitem [{\citenamefont {Gross}(1964{\natexlab{b}})}]{Gross:1964zz}%
  \BibitemOpen
  \bibfield  {author} {\bibinfo {author} {\bibfnamefont {F.}~\bibnamefont
  {Gross}},\ }\href {https://doi.org/10.1103/PhysRev.136.B140} {\bibfield
  {journal} {\bibinfo  {journal} {Phys. Rev.}\ }\textbf {\bibinfo {volume}
  {136}},\ \bibinfo {pages} {B140} (\bibinfo {year}
  {1964}{\natexlab{b}})}\BibitemShut {NoStop}%
\bibitem [{\citenamefont {Nystr{\"o}m}(1930)}]{Nystrom:1930aa}%
  \BibitemOpen
  \bibfield  {author} {\bibinfo {author} {\bibfnamefont {E.~J.}\ \bibnamefont
  {Nystr{\"o}m}},\ }\href {https://doi.org/10.1007/BF02547521} {\bibfield
  {journal} {\bibinfo  {journal} {Acta Mathematica}\ }\textbf {\bibinfo
  {volume} {54}},\ \bibinfo {pages} {185} (\bibinfo {year} {1930})},\ \bibinfo
  {note} {publisher: Institut Mittag-Leffler}\BibitemShut {NoStop}%
\bibitem [{\citenamefont {Atkinson}(1997)}]{Atkinson:1997fv}%
  \BibitemOpen
  \bibfield  {author} {\bibinfo {author} {\bibfnamefont {K.~E.}\ \bibnamefont
  {Atkinson}},\ }\href@noop {} {\emph {\bibinfo {title} {The Numerical Solution
  of Integral Equations of he Second Kind}}}\ (\bibinfo  {publisher} {Cambridge
  University Press},\ \bibinfo {year} {1997})\BibitemShut {NoStop}%
\bibitem [{\citenamefont {Kwon}\ and\ \citenamefont
  {Tabakin}(1978)}]{Kwon:1978}%
  \BibitemOpen
  \bibfield  {author} {\bibinfo {author} {\bibfnamefont {Y.~R.}\ \bibnamefont
  {Kwon}}\ and\ \bibinfo {author} {\bibfnamefont {F.}~\bibnamefont {Tabakin}},\
  }\href {https://doi.org/10.1103/PhysRevC.18.932} {\bibfield  {journal}
  {\bibinfo  {journal} {Phys. Rev. C}\ }\textbf {\bibinfo {volume} {18}},\
  \bibinfo {pages} {932} (\bibinfo {year} {1978})}\BibitemShut {NoStop}%
\bibitem [{\citenamefont {Andreev}(2017)}]{Andreev:2017kie}%
  \BibitemOpen
  \bibfield  {author} {\bibinfo {author} {\bibfnamefont {V.~V.}\ \bibnamefont
  {Andreev}},\ }\href {https://doi.org/10.1134/S1547477117010034} {\bibfield
  {journal} {\bibinfo  {journal} {Phys. Part. Nucl. Lett.}\ }\textbf {\bibinfo
  {volume} {14}},\ \bibinfo {pages} {66} (\bibinfo {year} {2017})}\BibitemShut
  {NoStop}%
\bibitem [{\citenamefont {Chen}(2012)}]{Chen:2012sv}%
  \BibitemOpen
  \bibfield  {author} {\bibinfo {author} {\bibfnamefont {J.-K.}\ \bibnamefont
  {Chen}},\ }\href {https://doi.org/10.1103/PhysRevD.86.036013} {\bibfield
  {journal} {\bibinfo  {journal} {Phys. Rev. D}\ }\textbf {\bibinfo {volume}
  {86}},\ \bibinfo {pages} {036013} (\bibinfo {year} {2012})},\ \bibinfo {note}
  {[Erratum: Phys.Rev.D 89, 099904 (2014)]}\BibitemShut {NoStop}%
\bibitem [{\citenamefont {Chen}(2013)}]{Chen:2013hna}%
  \BibitemOpen
  \bibfield  {author} {\bibinfo {author} {\bibfnamefont {J.-K.}\ \bibnamefont
  {Chen}},\ }\href {https://doi.org/10.1103/PhysRevD.88.076006} {\bibfield
  {journal} {\bibinfo  {journal} {Phys. Rev. D}\ }\textbf {\bibinfo {volume}
  {88}},\ \bibinfo {pages} {076006} (\bibinfo {year} {2013})}\BibitemShut
  {NoStop}%
\bibitem [{\citenamefont {Eyre}\ and\ \citenamefont {Vary}(1986)}]{Eyre:1986}%
  \BibitemOpen
  \bibfield  {author} {\bibinfo {author} {\bibfnamefont {D.}~\bibnamefont
  {Eyre}}\ and\ \bibinfo {author} {\bibfnamefont {J.~P.}\ \bibnamefont
  {Vary}},\ }\href {https://doi.org/10.1103/PhysRevD.34.3467} {\bibfield
  {journal} {\bibinfo  {journal} {Phys. Rev. D}\ }\textbf {\bibinfo {volume}
  {34}},\ \bibinfo {pages} {3467} (\bibinfo {year} {1986})}\BibitemShut
  {NoStop}%
\bibitem [{\citenamefont {Norbury}\ \emph {et~al.}(1992)\citenamefont
  {Norbury}, \citenamefont {Kahana},\ and\ \citenamefont
  {Maung~Maung}}]{Norbury:1992jv}%
  \BibitemOpen
  \bibfield  {author} {\bibinfo {author} {\bibfnamefont {J.~W.}\ \bibnamefont
  {Norbury}}, \bibinfo {author} {\bibfnamefont {D.~E.}\ \bibnamefont
  {Kahana}},\ and\ \bibinfo {author} {\bibfnamefont {K.}~\bibnamefont
  {Maung~Maung}},\ }\href {https://doi.org/10.1139/p92-009} {\bibfield
  {journal} {\bibinfo  {journal} {Can. J. Phys.}\ }\textbf {\bibinfo {volume}
  {70}},\ \bibinfo {pages} {86} (\bibinfo {year} {1992})}\BibitemShut {NoStop}%
\bibitem [{\citenamefont {Maung}\ \emph {et~al.}(1993)\citenamefont {Maung},
  \citenamefont {Kahana},\ and\ \citenamefont {Norbury}}]{Maung:1993aa}%
  \BibitemOpen
  \bibfield  {author} {\bibinfo {author} {\bibfnamefont {K.~M.}\ \bibnamefont
  {Maung}}, \bibinfo {author} {\bibfnamefont {D.~E.}\ \bibnamefont {Kahana}},\
  and\ \bibinfo {author} {\bibfnamefont {J.~W.}\ \bibnamefont {Norbury}},\
  }\href {https://doi.org/10.1103/PhysRevD.47.1182} {\bibfield  {journal}
  {\bibinfo  {journal} {Phys. Rev. D}\ }\textbf {\bibinfo {volume} {47}},\
  \bibinfo {pages} {1182} (\bibinfo {year} {1993})}\BibitemShut {NoStop}%
\bibitem [{\citenamefont {Hersbach}(1993)}]{Hersbach:1993xz}%
  \BibitemOpen
  \bibfield  {author} {\bibinfo {author} {\bibfnamefont {H.}~\bibnamefont
  {Hersbach}},\ }\href {https://doi.org/10.1103/PhysRevD.47.3027} {\bibfield
  {journal} {\bibinfo  {journal} {Phys. Rev. D}\ }\textbf {\bibinfo {volume}
  {47}},\ \bibinfo {pages} {3027} (\bibinfo {year} {1993})}\BibitemShut
  {NoStop}%
\bibitem [{\citenamefont {Tang}\ and\ \citenamefont
  {Norbury}(2001)}]{Tang:2001ii}%
  \BibitemOpen
  \bibfield  {author} {\bibinfo {author} {\bibfnamefont {A.}~\bibnamefont
  {Tang}}\ and\ \bibinfo {author} {\bibfnamefont {J.~W.}\ \bibnamefont
  {Norbury}},\ }\href {https://doi.org/10.1103/PhysRevE.63.066703} {\bibfield
  {journal} {\bibinfo  {journal} {Phys. Rev. E}\ }\textbf {\bibinfo {volume}
  {63}},\ \bibinfo {pages} {066703} (\bibinfo {year} {2001})},\ \Eprint
  {https://arxiv.org/abs/hep-ph/0103276} {arXiv:hep-ph/0103276} \BibitemShut
  {NoStop}%
\bibitem [{\citenamefont {Deloff}(2007)}]{Deloff:2007}%
  \BibitemOpen
  \bibfield  {author} {\bibinfo {author} {\bibfnamefont {A.}~\bibnamefont
  {Deloff}},\ }\href
  {https://doi.org/http://dx.doi.org/10.1016/j.aop.2006.10.004} {\bibfield
  {journal} {\bibinfo  {journal} {Annals of Physics}\ }\textbf {\bibinfo
  {volume} {322}},\ \bibinfo {pages} {2315 } (\bibinfo {year}
  {2007})}\BibitemShut {NoStop}%
\bibitem [{\citenamefont {Leit\~ao}\ \emph {et~al.}(2014)\citenamefont
  {Leit\~ao}, \citenamefont {Stadler}, \citenamefont {Pe\~na},\ and\
  \citenamefont {Biernat}}]{Leitao:2014jha}%
  \BibitemOpen
  \bibfield  {author} {\bibinfo {author} {\bibfnamefont {S.}~\bibnamefont
  {Leit\~ao}}, \bibinfo {author} {\bibfnamefont {A.}~\bibnamefont {Stadler}},
  \bibinfo {author} {\bibfnamefont {M.~T.}\ \bibnamefont {Pe\~na}},\ and\
  \bibinfo {author} {\bibfnamefont {E.~P.}\ \bibnamefont {Biernat}},\ }\href
  {https://doi.org/10.1103/PhysRevD.90.096003} {\bibfield  {journal} {\bibinfo
  {journal} {Phys. Rev. D}\ }\textbf {\bibinfo {volume} {90}},\ \bibinfo
  {pages} {096003} (\bibinfo {year} {2014})},\ \Eprint
  {https://arxiv.org/abs/1408.1834} {arXiv:1408.1834 [hep-ph]} \BibitemShut
  {NoStop}%
\bibitem [{Lan()}]{Lande+Kwon}%
  \BibitemOpen
  \href@noop {} {\bibinfo {title} {{A.\ Land\'e, as quoted in Ref.\
  }\cite{Kwon:1978}}}\BibitemShut {NoStop}%
\bibitem [{\citenamefont {Norbury}\ \emph {et~al.}(1994)\citenamefont
  {Norbury}, \citenamefont {Maung},\ and\ \citenamefont
  {Kahana}}]{Norbury:1994bh}%
  \BibitemOpen
  \bibfield  {author} {\bibinfo {author} {\bibfnamefont {J.~W.}\ \bibnamefont
  {Norbury}}, \bibinfo {author} {\bibfnamefont {K.~M.}\ \bibnamefont {Maung}},\
  and\ \bibinfo {author} {\bibfnamefont {D.~E.}\ \bibnamefont {Kahana}},\
  }\href {https://doi.org/10.1103/PhysRevA.50.2075} {\bibfield  {journal}
  {\bibinfo  {journal} {Phys. Rev. A}\ }\textbf {\bibinfo {volume} {50}},\
  \bibinfo {pages} {2075} (\bibinfo {year} {1994})}\BibitemShut {NoStop}%
\bibitem [{sta()}]{stadler_2024_14217822}%
  \BibitemOpen
  \href {https://doi.org/10.5281/zenodo.14217823} {\bibinfo {title}
  {https://doi.org/10.5281/zenodo.14217822}}\BibitemShut {NoStop}%
\end{thebibliography}%

\end{document}